\documentclass[3p,times,twocolumn]{elsarticle}
 \pdfoutput=1
\usepackage{graphicx}
\makeatletter
\def\ps@pprintTitle{%
 \let\@oddhead\@empty
 \let\@evenhead\@empty
 \def\@oddfoot{}%
 \let\@evenfoot\@oddfoot}
\makeatother

\usepackage[binary-units=true]{siunitx}
\usepackage{siunitx}
\usepackage{color}
\usepackage{verbatim}
\usepackage {longtable}
\usepackage {caption}
\usepackage{subcaption}
\usepackage {hhline}
\usepackage{soul}
\usepackage{amssymb}
\usepackage{lineno}
\usepackage{graphicx}
\usepackage{tikz}

\newcommand{\tikzcircle}[2][magenta,fill=magenta]{\tikz[baseline=-0.5ex]\draw[#1,radius=#2] (0,0) circle ;}%
\usepackage{amsmath}
\usepackage[misc]{ifsym}

\begin{document}
\begin{frontmatter}
\title{The evolution of the temperature field during cavity collapse in liquid nitromethane. Part I: Inert case}

\author[cav]{L.˜Michael}
\ead{lm355@cam.ac.uk}

\author[cav]{N.~Nikiforakis}
\ead{nn10005@cam.ac.uk}

\address[cav]{Laboratory for Scientific Computing, Cavendish Laboratory, Department of Physics, University of Cambridge, UK}


\begin{abstract}

This work is concerned with the effect of
cavity collapse in non-ideal explosives as a means of
controlling their sensitivity. The main objective is to
understand the origin of localised temperature peaks
(hot spots) which play a leading order role at the
early stages of ignition. To this end we perform two- and
three-dimensional numerical simulations of shock induced
single gas-cavity collapse in liquid nitromethane. Ignition
is the result of a complex interplay between fluid dynamics
and exothermic chemical reaction. In order to understand the
relative contribution between these two processes, we
consider in this first part of the work the evolution of
the physical system in the absence of chemical reactions.
We employ a multi-phase mathematical formulation which can
account for the large density difference across the gas-liquid
material interface without generating spurious temperature
peaks. The mathematical and physical models are validated
against experimental, analytic and numerical data. Previous
inert studies have identified the impact of the upwind (relative to
the direction of the incident shock wave) side of the
cavity wall to the downwind one as the main reason for the
generation of a hot-spot outside of the cavity, something which is also
observed in this work. However, it is also apparent that the
topology of the temperature field is more complex than
previously thought and additional hot spots locations exist,
which arise from the generation of Mach stems rather than
jet impact. To explain the generation mechanisms and
topology of the hot spots we carefully follow the complex
wave patterns generated in the collapse process
and identify specifically the temperature elevation or
reduction generated by each wave. This allows to track
each hot spot back to its origins. It is shown that the highest
hot spot temperatures can be more than twice
the post-incident shock temperature of the
neat material and can thus lead to ignition. By comparing
two-dimensional and three-dimensional simulation
results in the context of the maximum temperature observed
in the domain, it is apparent that three-dimensional calculations
are necessary in order to avoid belated ignition times in reactive
scenarios.
\end{abstract}

\begin{keyword}condensed phase explosives \sep cavity collapse \sep
temperature \sep nitromethane \sep hot spots \end{keyword}

\end{frontmatter}
\section{Introduction}

Many explosives used in mining are insensitive and do not detonate under typical shock loading conditions, to ensure their safe transportation to mining sites. Before being detonated in blastholes these explosives have to be sensitised by including artificial impurities such as gas cavities or glass micro-balloons.

The mechanism by which the sensitivity of the explosives is increased is a combination of hydrodynamic, thermodynamic and chemical effects when these cavities collapse under the influence of an incident shock wave. Upon collapse, regions of locally high pressure and high temperature are generated, which are often referred to as hot spots. If these regions are large enough and maintain high temperatures for a sufficient amount of time they can lead to the local ignition of the material and subsequently thermal runaway. In such cases, the time to ignition is observed to be considerably shorter than the time needed for the ignition of the neat explosive, thus the cavities are considered to sensitise the material. 

To this end, experimental and numerical studies have been performed, aiming to better understand  the physical processes behind the shock-induced cavity collapse and the sensitisation mechanism, which are imperative to the optimisation of the performance of non-ideal, inhomogeneous explosives.
Studying inhomogeneous explosives at the macro-scale, for example in the form of a rate stick simulation or experiment, provides insight to the overall effect the inhomogeneities have on the behaviour of the bulk of the explosive. However, the length-scales and time-scales governing the collapse of the cavities and the initiation and transition to detonation of the explosive bulk are massively disparate. As a result, it is very difficult for a macro-scale simulation or an experiment to simultaneously and adequately capture both processes. 

Hence, a lot of focus has turned to understanding the meso-scale physics by means of resolved simulations or experiments of cavity collapse to thereafter upscale to bulk. To reduce the complexity of the process induced by the chemical reactions, several researchers concentrated on the shock-induced cavity collapse phenomenon in isolation, as a purely hydrodynamical process, emphasising on the mechanical and the induced thermodynamic effects. 

For example, extensive experimental studies have been performed to identify the process governing the cavity collapse and determine the mechanical effects behind hot spot generation. A series of experiments of  air cavities (diameter $5-\SI{8}{\milli \meter}$) collapsing under the influence of shocks of various strengths ($0.3-\SI{5}{\giga \pascal}$) in gelatinous materials have been undertaken by Field and co-workers \cite{dear1988study,bourne1992shock}.
Bourne and Field \cite{bourne1992shock} studied the collapse process of isolated cavities of various shapes (circular, triangular, semi-circular, elliptical) in an inert medium. Dear and Field \cite{dear1988study} considered multi-cavity systems collapsing in an inert material and observed phenomena like jet deviation, asymmetric collapse and double jetting in columns of cavities and shielding effects and pressure amplification in rows of cavities. More recently, Swantek and Austin \cite{swantek2010collapse,swantek2010effect} studied collapsing cavity arrays under stress loading and analysed the velocity fields around them. 

Similar studies were done by numerical means, looking at the shock-cavity interaction process in various configurations. A lot of numerical studies include collapse in gaseous media or focus on pressure fields alone. Here, we discuss selectively only a small number of papers that involve non-gaseous media and concentrate on the ones demonstrating results on temperature fields, as these are directly relevant to this work. Mader \cite{mader1979numerical} was one of the first to simulate the collapse of a cavity in nitromethane and calculate the nitromethane temperature field around the cavity, to compare against the post-shock, neat material temperature. However, this work was limited by resolution, as the computers at the time were not yet capable of handling the heavy computation needed for adequately resolving such processes. Bourne and Milne \cite{bourne2002cavity,bourne2003temperature} modelled the temperature within the collapsing cavity, for circular, triangular and hemispherical isolated cavities and cavities in arrays. 
Ball et al.\ \cite{ball2000shock} and Hawnker and Ventikos \cite{hawker2012interaction} simulated a single cavity collapse in water and computed the water temperature field, while Lauer et al.\ \cite{lauer2012numerical}, Michael et al.\ \cite{michael2010numerical} and Betney et al.\ \cite{betney2015computational}  simulated the collapse of arrays in water and focused on the effect of the collapse on the pressure field and subsequent pressure amplification. 
Similarly, the collapse of circular and ellipsoidal cavities collapsing in an inert, hydrodynamically-modelled solid material was studied by Ozlem et al.\ \cite{ozlem2012numerical} who also focused on the pressure fields recovery and Kapila et al.\ \cite{kapila2015numerical} who presented temperature fields in an exemplary explosive. Limited results on the shock-cavity collapse in nitromethane were presented by Michael and Nikiforakis \cite{michael2015DetSymp}, in HNS by  Kittel and Yarrington \cite{kittell2016physically} and in HMX by Menikoff \cite{menikoff2004pore}. In an elastoplastic framework, Tran and Udaykumar \cite{tran2006simulation,tran2006simulation2} and Kapahi and Udaykumar \cite{kapahi2013dynamics} studied the response of inert and reactive HMX with micron-sized cavities, while Rai et al.\ \cite{rai2017high,rai2017collapse} considered the resolution required for reactive cavity collapse simulations and the sensitivity behaviour of elongated cavities in HMX.

The studies mentioned above have contributed to our understanding of the cavity-collapse mechanisms. However, there are still challenges in performing a complete, physical simulation of the initiation of a condensed phase explosive by shock-induced cavity collapse. An important step forward would be to address and rectify the current simulation challenges and work towards a complete simulation of the phenomenon. Such challenges include the non-trivial use of complex equations of state for describing the materials involved in the simulation, retaining at least 1000:1 density difference across the cavity boundary, maintaining oscillation free interfaces (in terms of pressures, velocities and temperatures), obtaining realistic temperature fields in the explosive matrix and heavy computations required for well-resolved, three-dimensional simulations.

A major challenge worthy of special attention is the accurate and oscillation-free recovery of temperature fields in the explosive matrix. This is of critical importance as the ignition process of the energetic material is a temperature-driven effect, thus the accurate prediction of ignition relies on physically meaningful temperatures. 
An essential prerequisite for recovering accurate temperature fields is the characterisation of the explosive material by means of a realistic equation of state, for example in Mie-Gr\"uneisen rather than in polytropic form, which increases the complexity of the underlying governing equations and hence the difficulty of their numerical integration. Moreover, the large density difference across the material interface (at least 1000:1 for a gas bubble in a liquid) generates additional problems, compared to a uniform material simulation. Assuming that a non-oscillatory, interface capturing numerical scheme is employed, thin but purely numerical mixture zones are generated at the material boundaries, which can give rise to velocity, pressure and temperature oscillations \cite{banks2007high,kapila2007study,michael2016hybrid}. Even if the velocity and pressure oscillations are eliminated  (e.g.\ by an energy correction step \cite{banks2007high,michael2016hybrid} or by augmenting the Euler system of equations with equations evolving parameters of the equation of state \cite{shyue1998efficient,shyue1999fluid,shyue2001fluid,wang2004thermodynamically}), the temperature oscillations may persist, as demonstrated in previous work by the authors (Fig.\ 4 in  \cite{michael2016hybrid}). 
This could be attributed to the fact that the specific heat of the artificial mixture cannot be physically derived and hence must be constructed mathematically. Independent of their origin, the temperature oscillations would be destructive in any ignition studies, as they would spuriously activate the temperature-dependent reaction rate, thus generating false ignition. It is, thus, important to ensure that spurious numerical artefacts are either removed or do not arise in the first place.

Most work found in the literature that present temperature fields is either focusing on the temperature inside the cavity only, done in two-dimensions or use idealised equations of state. In this work, we overcome the difficulties in numerically simulating the cavity collapse mentioned above (including the spurious temperature oscillations) by exercising a mathematical model proposed by the authors in previous work \cite{michael2016hybrid}, allowing for a complete shock-cavity interaction simulation.  We simulate the three-dimensional collapse of isolated air cavities in nitromethane, using the Cochran-Chan equation of state for the explosive and demonstrating the validity of this equation of state in the regime of pressures found in the simulations by comparing to experiment. The temperature fields recovered are also free from spurious oscillations and the 1000:1 density difference between the cavity and ambient material is captured directly. In the first instance we consider the induced temperature field in the explosive in the absence of chemical reactions so that hydrodynamical effects are elucidated. However, the oscillation-free temperature fields will allow for the use of a temperature-dependent reaction rate law in future work. In this work, we move beyond solely reporting potential hot spot loci and identifying high temperature peaks; we look in depth how the mechanical effects (e.g.\ the generation and propagation of waves) lead to the thermodynamical effects (e.g.\ elevation of temperatures) that generate the localised hot spots. Each wave generated in the complex cavity collapse scenario is labelled and its effect on the temperature field is thereafter determined. In experiments such detail of study is limited by instrumentation, the locus of luminescence downstream the cavity is noted as the main ignition site and the jet impact as the principal ignition mechanism (e.g.\ \cite{bourne1992shock,bourne1991bubble,bourne1999explosive}). In numerical studies, where more detailed wave patterns can be obtained, the shock waves produced upon jet impact are considered as a single wave moving outwards of the cavity remains (e.g. \cite{bourne2002cavity}). We demonstrate, however, that two distinct shock wave components can be identified upon collapse; one travelling downstream and one travelling upstream. It is also demonstrated that it is the upstream travelling wave that leads to the highest temperature locus in collapse scenarios of the incident shock pressure considered here. We identify the regions with temperatures that are higher than the post-shock temperature and are candidates for ignition loci in reactive simulations. In this process, we report a new hot spot location, in the Mach stem generated at late stages of the collapse. This hot spot is omitted in previous studies  (e.g. \cite{bourne2002cavity}) but is considered here in detail as its effect could be similar to that of the principal hot spot along the cavity centreline.  By comparing to the equivalent two-dimensional cavity collapse we demonstrate also the importance of performing this study in a three-dimensional framework as the two-dimensional simulations are shown to under-predict the temperature fields in some stages and over-predict them in others, something that could lead to no or false ignition sites.

 In Part II of this work, that will follow in the future, the reactions in nitromethane will be activated and the ignition of the material can then be traced back to the elevation of the temperature fields and accountable waves described here.

The rest of the paper is  outlined as follows; the next section presents the underlying formulation used in this work in terms of governing PDEs, the equations of state that close the system and the method for temperature recovery. A section on validation follows, where the mathematical system and numerical methods are tested against known solutions. The mathematical and physical models are also validated against theoretical and experimental temperatures of shocked nitromethane.
In the results section that follows we present the three-dimensional collapse of an air cavity in inert, liquid nitromethane and follow the events leading to the generation of locally high temperatures, study the temperature field topology and generated hot spots, the evolution of pressure and temperature fields on constant latitude lines and the maximum temperatures in the 2D and 3D simulations. Finally a section with the conclusions of this work is presented.

\section{Mathematical and physical model}
\begin{figure}[!tb]
\centering
\includegraphics[width=0.5\textwidth]{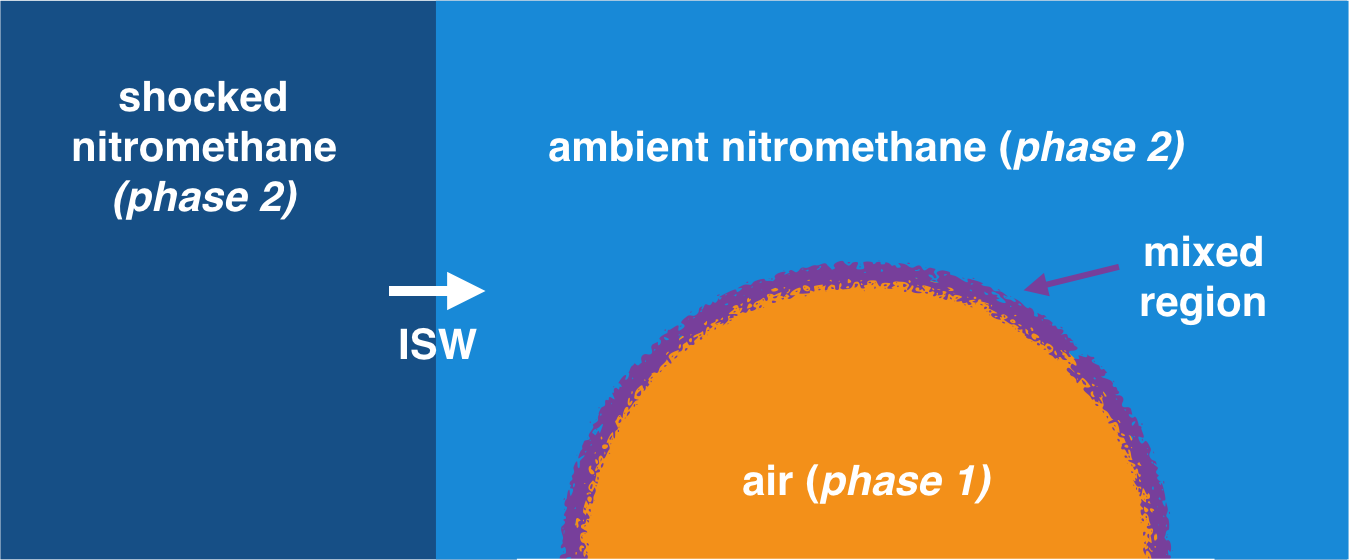}
\caption{Schematic of a two-dimensional projection of the simulations considered in this work. Phases 1 and 2 as described by the mathematical model are identified and the incident shock wave leading to the cavity collapse is indicated as ISW.}
\label{sketch}
 \end{figure}
A formulation which rectifies the issues commonly presented in shock-bubble simulations discussed in the previous section was proposed by Michael and Nikiforakis \cite{michael2016hybrid} (MiNi16). This formulation considers the cavity as an inert phase (\emph{phase 1}) and the surrounding material as a reacting phase (\emph{phase 2}). This work focuses on the collapse of cavities in an inert medium, hence we neglect the reaction source terms and the products of reaction and as a result we use a reduced form of the MiNi16 formulation. The reactive terms are reinstated in the follow-up, Part II of this paper. Considering the gas inside the cavity as \emph{phase 1} and the liquid nitromethane around the cavity as  \emph{phase 2}, as shown in Fig.\ \ref{sketch}, the governing equations for this system take the form
\begin{eqnarray}
\frac{\partial z_1 \rho_1}{\partial t}+\nabla \cdot(z_1 \rho_1 \mathbf{u})& = & 0, \nonumber \\
\frac{\partial z_2 \rho_2}{\partial t}+\nabla \cdot(z_2 \rho_2 \mathbf{u})& = & 0, \nonumber\\ \label{Hyb:system}
\frac{\partial}{\partial t}(\rho \mathbf{u_k})+\nabla \cdot(\rho \mathbf{u_ku})+\frac{\partial p}{\partial{\mathbf{x_k}}} & = & 0, \\ 
\frac{\partial}{\partial t}(\rho E)+\nabla \cdot [ (\rho E+p)\mathbf{u}] & = & 0, \nonumber \\ 
\frac{\partial z_1 }{\partial t} + \mathbf{u} \cdot \nabla z_1  & = & 0, \nonumber  
\end{eqnarray}
where for $i=1,2$, $\rho_i$  are the densities for the air and nitromethane, $z_i$ are their corresponding volume fractions 
($z_1+z_2=1$),
$\rho$ is the total density given by $\rho=z_1\rho_1+z_2\rho_2$, $\mathbf{u}$
is the velocity vector and $p$ is the total pressure. The total specific energy is given by $E=\frac{1}{2}u^2+e$, where $e$ is the total specific internal
energy.  
 The mixture rule for the total internal energy is given 
by $\rho e=\rho_1z_1e_1+\rho_2z_2e_2$, where $e_i$ for $i=1,2$ are the specific internal energies for the air and nitromethane, 
given by their corresponding equations of state. A mixture rule for $\xi=\frac{1}{\gamma-1}$, where $\gamma$ is the total 
adiabatic index, is also required and in this case is given by $\xi=z_1\xi_1+z_2\xi_2$. 
The sound speed for the total mixture is given by 
\begin{equation}
\xi c^2 = y_1\xi_1c_1^2+y_2\xi_2c_2^2,
\end{equation}
 where $y_i$ is the mass fraction of \emph{phase i}, given by $y_i=\frac{\rho_iz_i}{\rho}$, for $i=1,2$.

\subsection{Equations of state}
\label{Sec:EoS}
To close the system, the Cochran-Chan equation of state \cite{saurel2007relaxation} is employed to describe the liquid nitromethane. 
This is an equation of state of Mie-Gr\"uneisen form and is given by
\begin{equation}
\label{MG}
p(\rho,e)=p_{\text{ref}}(\rho)+\rho\Gamma(\rho)[e-e_{\text{ref}}(\rho)]
\end{equation}
with reference pressure given by 
\begin{equation}
\label{CCeos1}
p_{\text{ref}}(\rho)=\mathcal{A}\Big{(}\frac{\rho_0}{\rho}{\Big{)}^{-\mathcal{E}_1}-\mathcal{B}\Big{(}\frac{\rho_0}{\rho}{\Big{)}^{-\mathcal{E}_2}}},\end{equation}
reference energy given by
\begin{multline}
\label{CCeos2}
e_{\text{ref}}(\rho)=\frac{-\mathcal{A}}{\rho_0(1-\mathcal{E}_1)}\Big{[}\Big{(}\frac{\rho_0}{\rho}\Big{)}^{1-\mathcal{E}_1}-1\Big{]} \\ +\frac{\mathcal{B}}{\rho_0(1-\mathcal{E}_2)}\Big{[}\Big{(}\frac{\rho_0}{\rho}\Big{)}^{1-\mathcal{E}_2}-1\Big{]}
\end{multline}
and Gr\"uneisen coefficient
$\Gamma(\rho)=\Gamma_0.$
 gas inside the cavity is modelled by the ideal gas equation of state, which is of 
Mie-Gr\"uneisen form as well, with $p_{\text{ref}}=0$ and $e_{\text{ref}}=0$ . The parameters for the equations of state of the 
two materials are given in Table \ref{EoSNMAir}.

\begin{table*}[!t]
\centering
\caption{Equation of state parameters for  nitromethane and air.}
\label{EoSNMAir}
 \begin{tabular}{ccccccccc}
 \hline\noalign{\smallskip}
\footnotesize \textbf{Equation of state} & $\Gamma_0$ & $\mathcal{A}$  & $\mathcal{B}$   & $\mathcal{E}_1$  & $\mathcal{E}_2$ & $\rho_0$  & $c_v $  & Q  \\ 
\footnotesize \textbf{parameters} &-  &$\SI{}{\giga \pascal}$  &$\SI{}{\giga \pascal}$  &- & - &$\SI{}{\kilogram \per \meter \tothe{3}}$ &  $\SI{}{\joule \per \kilogram \per \kelvin}$ & $\SI{}{\mega \joule \per \kilogram }$ \\ 
\noalign{\smallskip}\hline\noalign{\smallskip}
\footnotesize Nitromethane & 1.19 & 0.819 & 1.51  & 4.53 &  1.42 &  1134 & 1714 & 4.48  \\
\footnotesize Air & 0.4 & - & -  & - & - &  - & 718 & 0  \\
\noalign{\smallskip}\hline
\end{tabular}
\end{table*}

\subsection{Recovery of temperature}

The multi-phase nature of the model allows for separate temperature fields to be computed for each material as
\begin{equation}
T_i=\frac{p-p_{\text{ref}_i}(\rho)}{\rho_i\Gamma_ic_{v_i}}, \quad \mbox{for} \quad  i=1,2.
\label{temp}
\end{equation} 
As a result, the nitromethane temperature ($T_{\text{NM}}=T_2$) is computed explicitly from the equation of state and can be used directly in the reaction rate law.

Computing the temperature of a general condensed phase explosive ($T_{\text{CF}}=T_2$) can involve completing the equation of state starting from the basic thermodynamic law $T \frac{dS}{dv}=c_v\frac{dT}{dv}+c_v\frac{\Gamma T}{v}$ and by integrating to obtain a reference temperature ($T_{{\text{ref}_\text{{CF}}}}$) such that:

\begin{equation}
T_{\text{CF}}=\frac{p-p_{{\text{ref}_\text{{CF}}}}(\rho)}{\rho_{\text{CF}}\Gamma_ic_{v_\text{CF}}} - T_{{\text{ref}_\text{{CF}}}}.
\label{tempref}
\end{equation} 

When the reference curve is an isentrope, $dS/dv=0$ and hence we can simply compute $T_{\text{ref}_\text{{CF}}}=T_0(\frac{\rho}{\rho_0})^\Gamma$. When the reference curve is a Hugoniot curve, the basic thermodynamic law cannot be integrated directly and often the Walsh Christian technique and numerical ode-integration techniques are used to compute the reference Hugoniot temperature. 

For this work, substitution of the parameters of the equation of state for nitromethane and imposing an initial temperature of $298K$ for $\rho=\rho_0$ gives $T_0=0$. This is in line with other work using the Cochran-Chan equation of state, where $T_0$ takes zero or very small values \cite{saurel1998riemann,massoni1999mechanistic,shukla2010interface,genetier2014effect}. The form (\ref{temp}) gives temperatures that match experiments as demonstrated in the validation section, but for other materials or other equations of state (e.g. shock Mie-Gr\"uneisen) care should be taken as a different reference curve (as per Eq.\ \ref{tempref}) would be necessary.

The ideal gas equation of state results in the overheating of the gas inside the cavity since high pressures are reached within the cavity during the collapse process. However, at these timescales heat transfer would not physically take place and thus the cavity temperature would not affect the ambient nitromethane temperature. Since temperatures inside the cavity are not of interest for this work, they are not presented henceforth.

Note that the two-phase nature of the model allows for large ($>$1000:1) density gradients to be sustained across material boundaries and both the density and temperature fields are maintained oscillation-free.

\section{Validation}
System (\ref{Hyb:system}) is integrated numerically using a high resolution shock-capturing numerical scheme, namely the 
MUSCL-Hancock finite volume method with an underlying HLLC Riemann solver. Hierarchical, structured, adaptive mesh refinement 
(AMR) is used to dynamically increase the resolution locally \cite{bates2007richtmyer}. 

The aim of this section is to validate the resulting code using test problems with known solutions and to assess the suitability of the physical models to predict realistic temperature fields.

\subsection{Air pocket collapsing in water}

The shock-induced collapse of an air bubble in water is used as a validation example for the mathematical and numerical methodologies in this work, due to its close relation to the application we are considering. We follow the set up considered, among others, by Terashima and Tryggrason \cite{terashima2010front} and Xu and Liu \cite{xu2017explicit}. As the solution to this problem is self-similar and to compare directly with results from both  \cite{terashima2010front} and \cite{xu2017explicit}, we consider this test in non-dimensional variables. Denoting the dimensional variables by over-bars, the non-dimensional variables in consideration are given as:

\begin{equation}
\rho=\frac{\bar{\rho}}{\rho_0}, \quad u=\frac{\bar{u}}{u_0}, \quad p=\frac{\bar{p}}{\rho_0u_0^2}, \quad x=\frac{\bar{x}}{l_0}, \quad t=\frac{\bar{t}}{l_0/u_0},
\end{equation}
where $\rho_0=\SI{1000}{\kilogram \per \meter \tothe{3}}$, $u_0=\SI{10}{\meter \per \second}$ and $l_0=\SI{1}{\milli \meter}$ for \cite{terashima2010front} and $l_0=\SI{1}{\meter}$ for \cite{xu2017explicit}.

A domain of size $(x,y)\in[-6,6]\times[0,6]$ is considered with base resolution $250\times125$ and three levels of AMR, doubling the resolution with each refinement level, leading to an effective resolution of $2000\times1000$ cells and an air bubble of initial diameter $D=6$. The initial conditions for this test are given  in non-dimensional form in Table \ref{ICval}.

The air is modelled as an ideal gas and the water as a stiffened gas. Both equations of state can be written in the Mie-Gr\"uneisen form (\ref{MG}), with $p_{\text{ref}}(\rho)=e_{\text{ref}}(\rho)=0$ and $\Gamma_0 = \gamma - 1 = 0.4$ for the air and  $p_{\text{ref}}(\rho)=6000$, $e_{\text{ref}}(\rho)=0$ and $\Gamma_0 = \gamma - 1 = 3.4$ for the water. In this work only the upper part of the bubble is simulated thus a reflective boundary condition is used at the bottom horizontal domain boundary. All other domain boundary conditions are taken to be transmissive.

\begin{table*}
\centering
\caption {Initial conditions for the shock-induced cavity collapse in  water considered for validation purposes}
\label{ICval}
\begin{tabular}{lccccccc}
\hline\noalign{\smallskip}
 & Material & $\rho_1$ & $\rho_{2}$ & $u$ & $v$ & $p$  & $z$  \\
\noalign{\smallskip}\hline\noalign{\smallskip}
$x<0.6$ & shocked water & $1.59\times10^{-3}$ & 1.325 & 68.0525 & 0.0 & 19153 & $10^{-6}$ \\
$x\ge 0.6$  & ambient water  & $1.2\times10^{-3}$ & 1.0 & 0.0 & 0.0 & 1.0 & $10^{-6}$ \\
 Bubble & air & $1.2\times10^{-3}$ & 1.0 & 0.0 & 0.0 & 1.0 & $1-10^{-6}$ \\
\noalign{\smallskip}\hline
\end{tabular}
\end{table*}

\begin{figure}
\begin{minipage}{\columnwidth}
\centering
        \begin{subfigure}[b]{0.49\textwidth}
\includegraphics[width=0.99\textwidth]{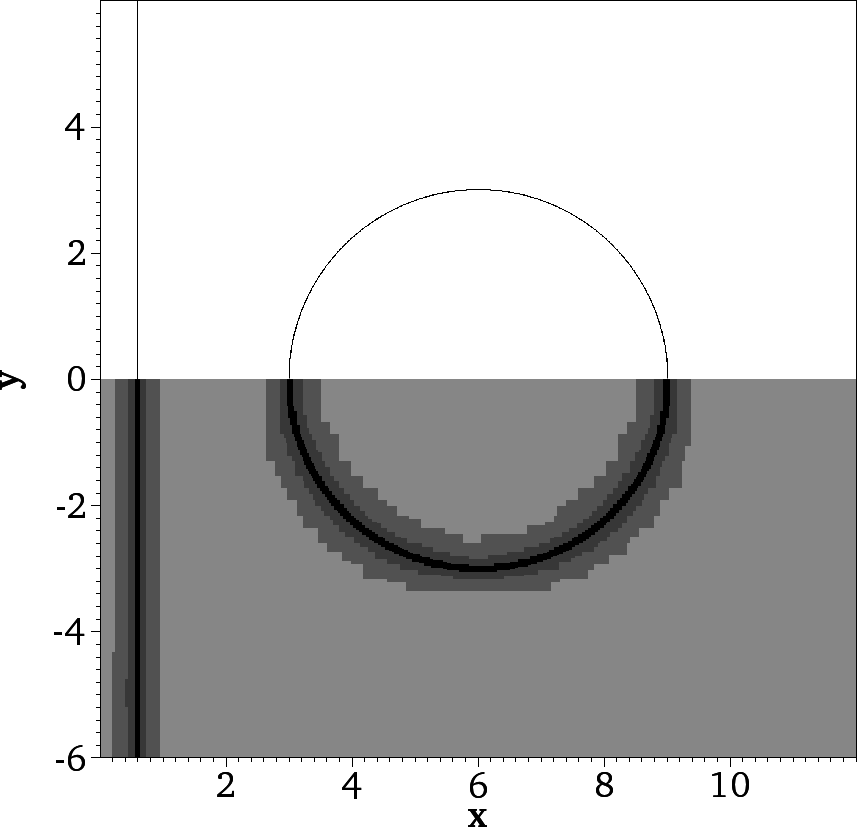}
        \end{subfigure}
\vspace{0.15cm}
\begin{subfigure}[b]{0.49\textwidth}
\includegraphics[width=0.99\textwidth]{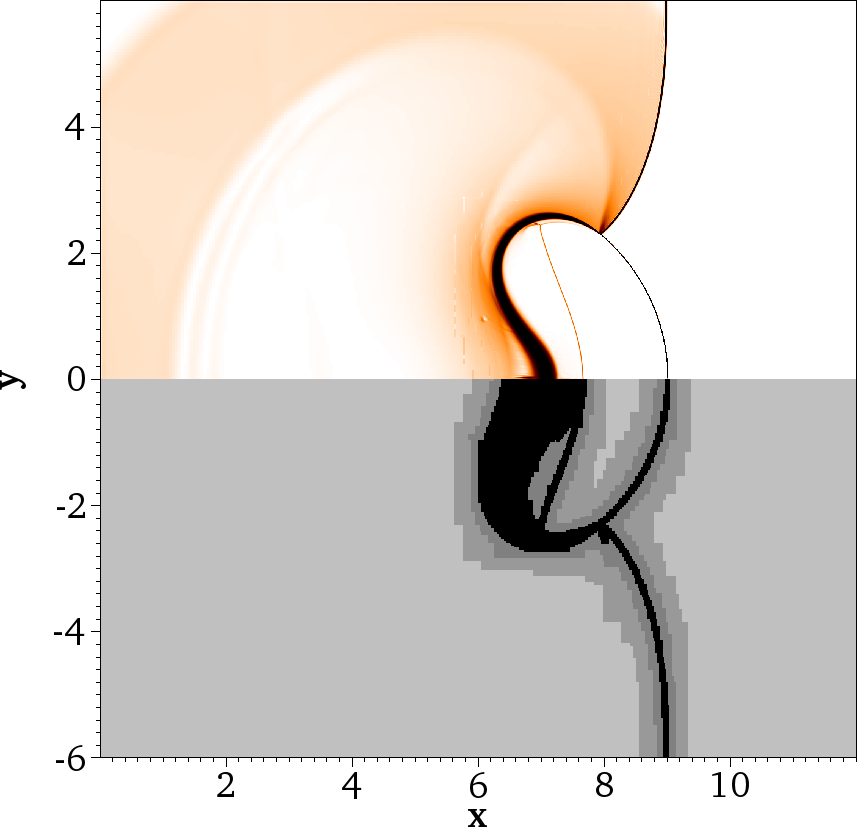}
        \end{subfigure}
\vspace{0.15cm}
\end{minipage}
\begin{minipage}{\columnwidth}
\centering        
\begin{subfigure}[b]{0.49\textwidth}
\includegraphics[width=0.99\textwidth]{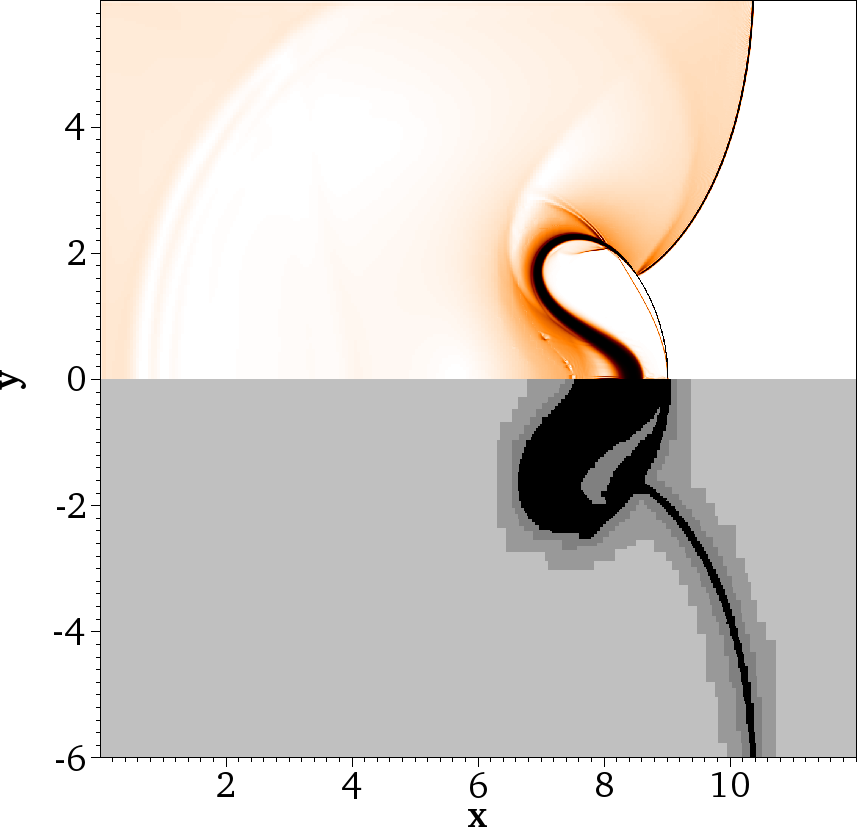}
        \end{subfigure}
\begin{subfigure}[b]{0.49\textwidth}
\includegraphics[width=0.99\textwidth]{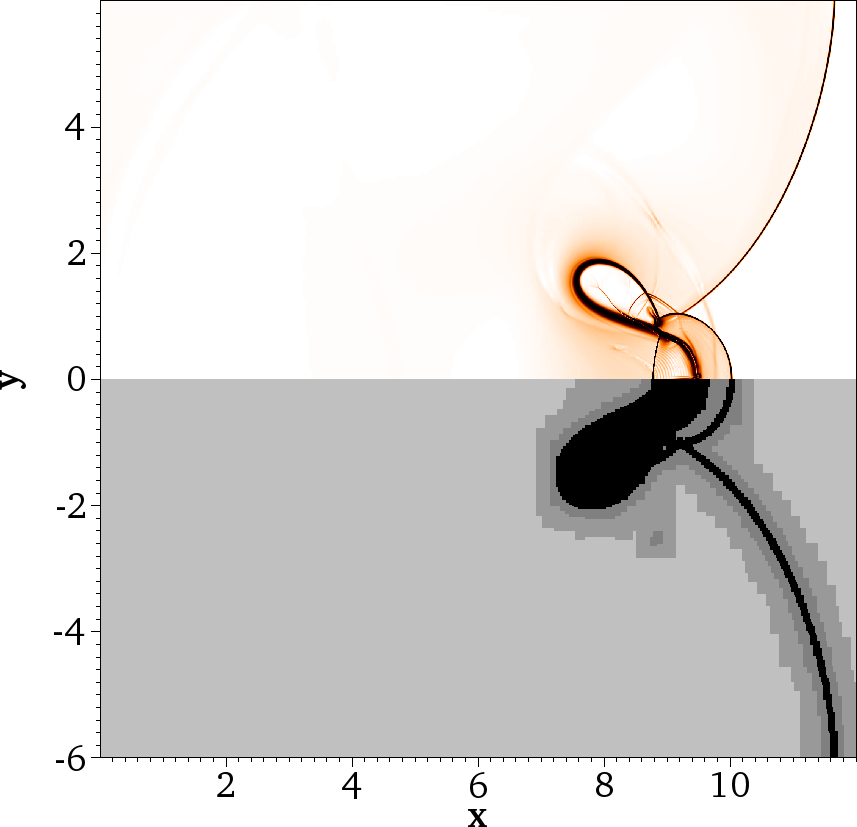}
        \end{subfigure}
\end{minipage}
\caption{The top half of each figure illustrates density mock-schlieren plots of the collapse process
of an air bubble in water, induced by the impact of a Mach 1.72 shock wave. The bottom half of each figure demonstrates the distribution of the AMR grids, with three levels of refinement. The figures show results at $t=0$ (top left), $t=3$ (top right), $t=3.5$ (bottom left), $t=4$ (bottom right).}
\label{waterShockBubble}
\end{figure}

The evolution of the bubble collapse is seen in Fig.\ \ref{waterShockBubble} where the density mock-schlieren is used to visualise the waves at the top half of the figures. On the bottom of each figure the AMR grids are shown reflected about the $x$-axis.

\begin{figure}
\centering
\begin{subfigure}[b]{0.4\textwidth}
\subcaptionbox[position=b]{The evolution of the bubble boundary due to its interaction with the incident shock wave, at instances $t=1,1.5,2,2.5,3,3.5,4$. The top half ($y>0$) shows results from this work and the bottom half ($y<0$) shows results from \cite{xu2017explicit}.\label{comparisonA}}{\includegraphics[width=\textwidth]{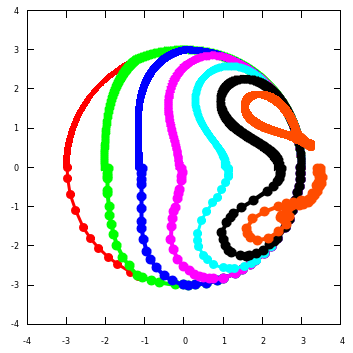}}
\end{subfigure}
\qquad
\begin{subfigure}[b]{0.47\textwidth}
\subcaptionbox[position=b]{Comparison of the evolved, non-dimensional height (H/D) and width (W/D) between this work and results from  \cite{terashima2010front} and \cite{xu2017explicit}.\label{comparisonB}}{\includegraphics[width=\textwidth]{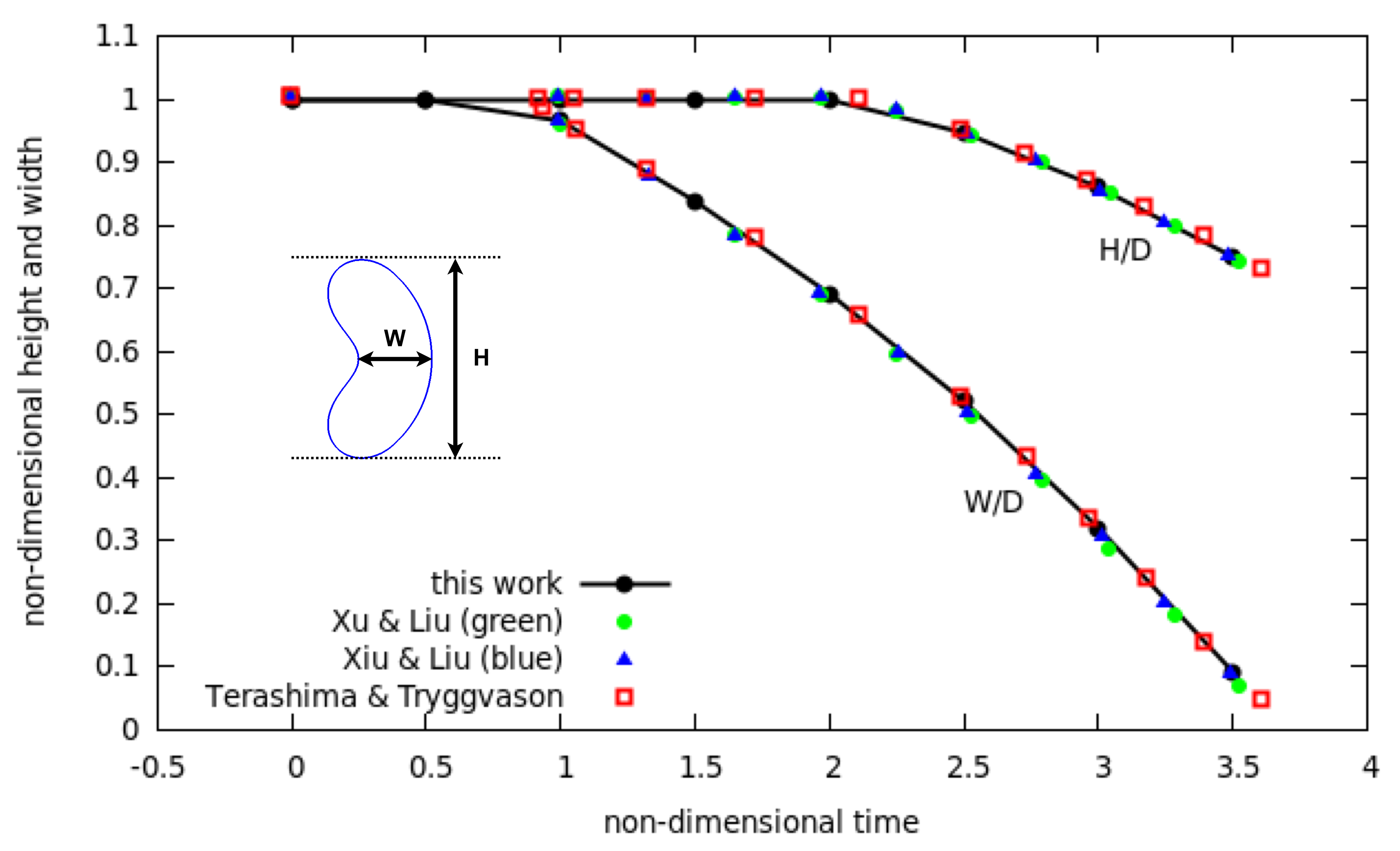}}
\end{subfigure}
\caption{Comparison of (a) the bubble boundary location and (b)  bubble dimensions from the results of this work and results found in literature.}
\label{comparison}
 \end{figure}

In Fig.\ \ref{comparisonA} the evolution of the bubble boundary under the effect of the incident shock wave is shown at time instances $t=1,1.5,2,2.5,3,3.5$ and 4. The top half of the image ($y>0$) shows results from this work and the bottom half ($y<0$) shows results from \cite{xu2017explicit}. The two sets of results match reasonably well and small differences are attributed to the fact that a diffused interface method (shock-capturing) is used in this work whereas a shock-tracking method is used in \cite{xu2017explicit}. In Fig.\  \ref{comparisonB} we present the evolved, non-dimensional, normalised with respect to the bubble diameter width and height of the bubble. We compare the width and height obtained from this work to the results of \cite{terashima2010front} and \cite{xu2017explicit}.

\subsection{Shocked nitromethane temperature}
\begin{figure}[!th]
\centering
\includegraphics[width=0.48\textwidth]{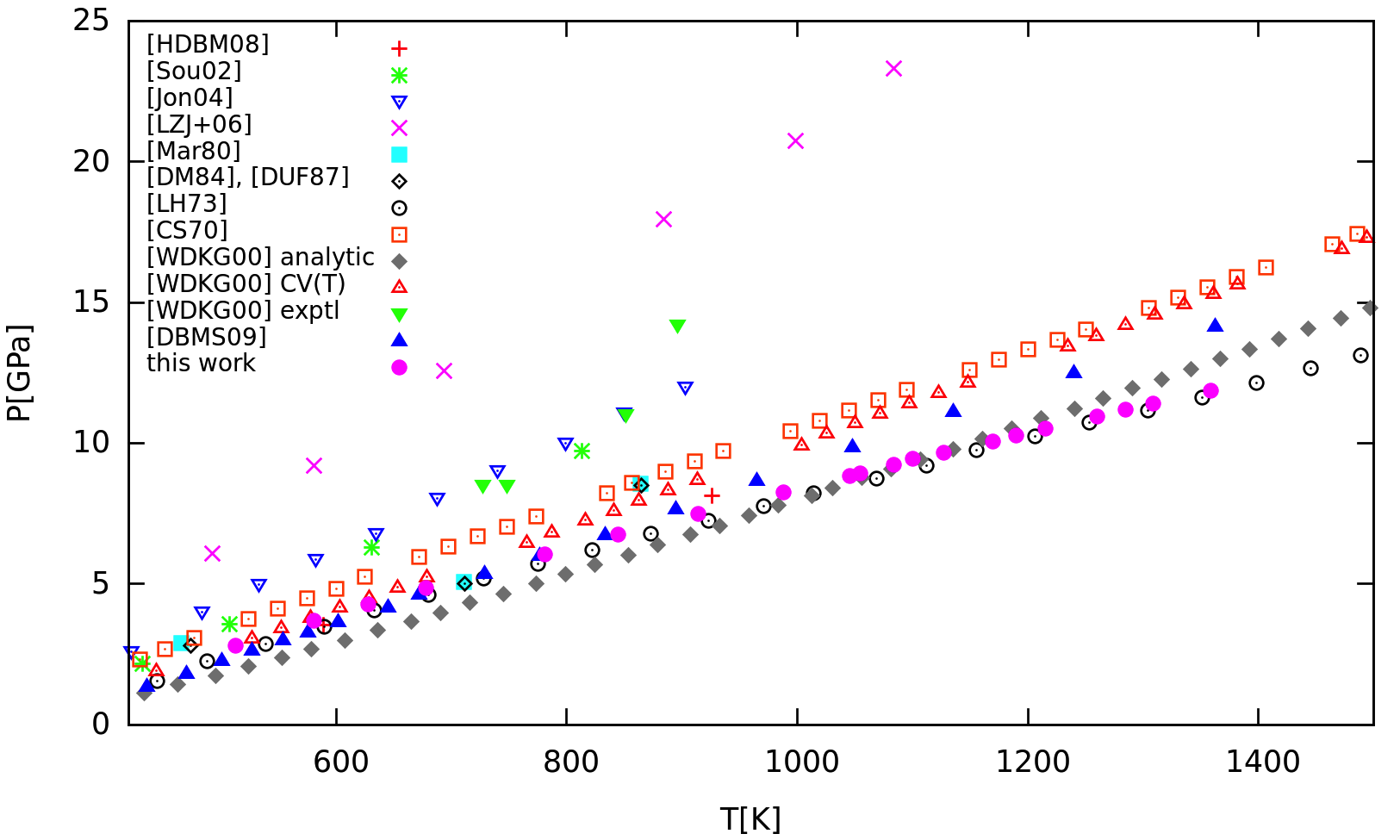}
\caption{Post-shock temperatures at different pressures. Comparison between results from this work (solid circles) and 
a compilation of published experimental, analytic and molecular simulation data.} \label{NMtemp}
\end{figure}

To validate the physical models used in this work we compare the temperature of the shocked, neat, non-reacting nitromethane computed from our simulations with 
experimentally measured values, as well as with the temperature calculated from analytic theory and molecular simulation, 
for various post-shock pressures; this compilation is presented in Fig.\ \ref{NMtemp}. 
The  set of experimental data most commonly used as a set of reference values
is given by Lysne and Hardesty \cite{lysne1973fundamental} and have been deduced numerically
from experimentally determined parameters. Other experiments measuring post-shock temperatures
include the work by Winey et al.\ \cite{winey2000equation}\ and Hervou{\"e}t et al.\ \cite{marsh1980lasl},
 as well as work by Delpuech and Menil \cite{delpuech1984raman} and Dufort \cite{dufort1987mesures}. 
Analytic work includes Winey et al.\ \cite{winey2000equation}\
 and Cowperthwaite and Shaw \cite{cowperthwaite1970c}, while results from microscopic simulations
are given by Soulard \cite{soulard2002shock}, Jones \cite{jones2004equation}, Liu et al.\ \cite{liu2006compressibility},
 Hervou{\"e}t et al.\ \cite{hervouet2008microscopic}\ and Desbiens et al.\ \cite{desbiens2009molecular}.

The solid circles (\tikzcircle{1.5pt}) in Fig.\ \ref{NMtemp} present the temperature
of neat nitromethane predicted from our simulations, calculated by 
expression (\ref{temp}). It is important to note the wide spread of the results from different sources and the fact that there is no indication  
of experimental error. Nevertheless,  the post-shock temperatures predicted from our simulations are well 
within the bounds of this compilation and they are closest  to the ones by Lysne and Hardesty \cite{lysne1973fundamental}. This ensures the validity of the temperature field for the explosive material, before the inclusion of cavities.

\section{The collapse of a single cavity in liquid non-reacting nitromethane}
\label{3Dresults}

In this section, an isolated air-filled cavity of radius $\SI{0.08}{\milli \meter}$\footnote{typical size of large cavities used in mining applications} collapsing in non-reactive liquid nitromethane due to a 10.98 GPa incident shock wave (ISW) is
considered. The air inside the cavity is modelled by the ideal gas equation of
state and the nitromethane is modelled by the Cochran-Chan equation of state as described in Sec.\ \ref{Sec:EoS}, with parameters as listed in Table \ref{EoSNMAir}. As the ignition and thermal runaway in an explosive are attributed to the complex interaction between non-linear gas-dynamics and chemistry, it is intuitive to consider in the first instance the induced temperature field in the explosive in the absence of chemical reactions. This will allow the purely gas-dynamical effects to be elucidated. The simulation is performed in three-dimensions, with effective grid size $dx=dy=dz=\SI{0.3125}{\micro \meter}$, 
selected after the convergence study presented in Appendix \ref{appendixA}. 
The initial conditions for the simulation are given in Table \ref{ICs}, corresponding to the sketch of Fig.\ \ref{sketch}.

\begin{table*}[!t]
\centering
\caption {Initial conditions for the shock-induced cavity collapse in inert nitromethane considered in this work}
\label{ICs}
\begin{tabular}{lcccccc}
\hline\noalign{\smallskip}
 & Material & $\rho_1$ & $\rho_{2}$ & $u$ & $p$  & $z$  \\
 & & [$\SI{}{\kilogram \per \meter \tothe{3}}$]& [$\SI{}{\kilogram \per \meter \tothe{3}}$] & [$\SI{}{\meter \per \second}$] & [$\SI{}{\pascal}$] &  \\
\noalign{\smallskip}\hline\noalign{\smallskip}
$x<100\SI{}{\micro \meter}$ & shocked nitromethane & 2.4 & 1934.0 & 2000.0 & $10.98\times10^{9}$ & $10^{-6}$ \\
$x\ge 100\SI{}{\micro \meter}$  & ambient nitromethane & 1.2 & 1134.0 & 0.0 & $1\times10^{5}$ & $10^{-6}$ \\
 Bubble & air &1.2 & 1134.0 & 0.0 &  $1\times10^{5}$ & $1-10^{-6}$ \\
\noalign{\smallskip}\hline
\end{tabular}
\end{table*}

The aim of this section is to follow in detail the generation of complex wave pattern during the collapse process and identify the effect of each wave on the nitromethane temperature field. This will lead to identifying which waves are generating the high-temperature regions in the explosive and the magnitude of the temperature field that could account for the local ignition in a reactive simulation.

\subsection{Early stages of the three-dimensional cavity collapse}

\begin{figure*}
        \begin{subfigure}[b]{0.49\textwidth}
\includegraphics[width=0.99\textwidth]{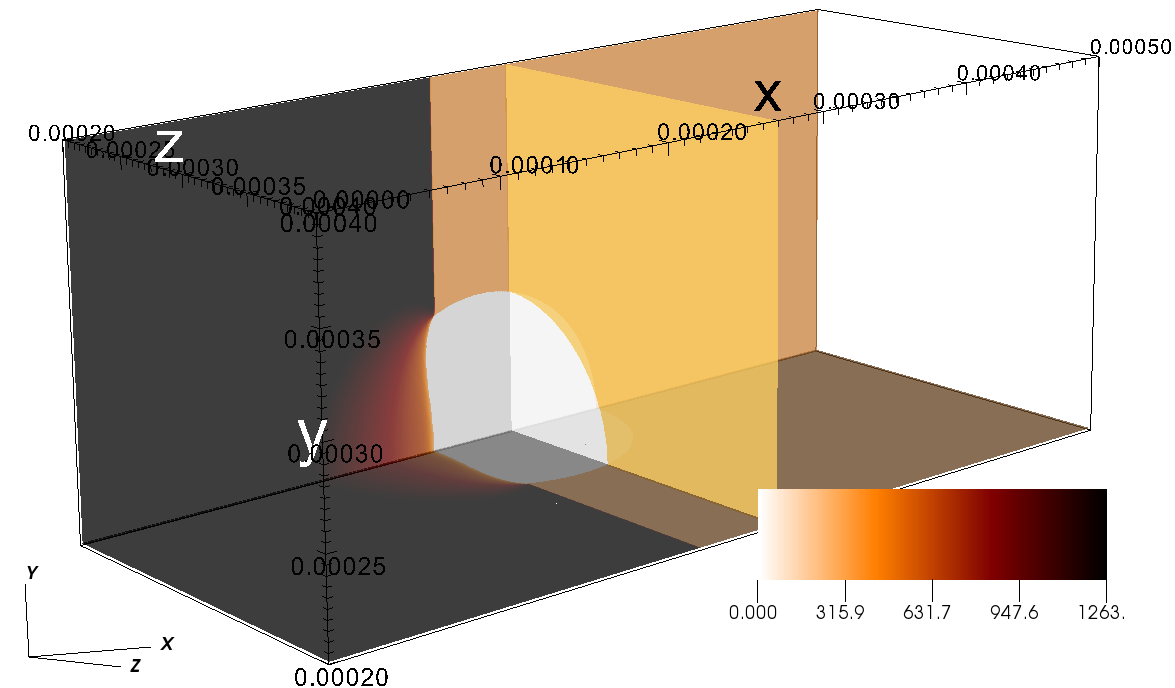}
        \end{subfigure}
        \begin{subfigure}[b]{0.49\textwidth}
\includegraphics[width=0.99\textwidth]{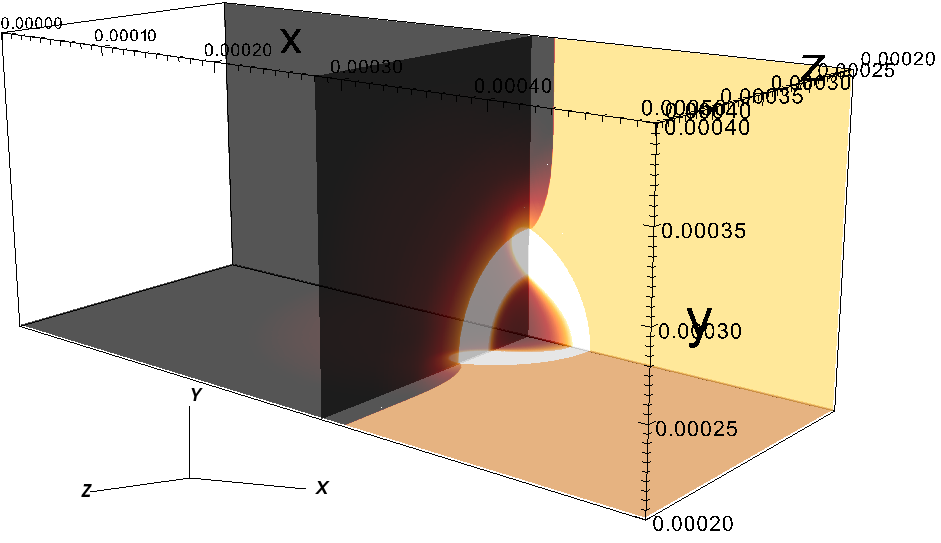}
        \end{subfigure}
\caption{ A sample of plots to visualise the shock-bubble collapse in 3D. Slices through the centre of the cavity in the $xy$ and $xz$ planes are shown, along with a selected slice on the $yz$ plane. Early stages of the collapse are shown, illustrating the generation of the release wave in nitromethane and the lobe generation before jet impact.  All axes in \SI{}{\meter}.}
\label{3Dtemp}
\end{figure*}

\begin{figure*}
\begin{minipage}{2\columnwidth}
\centering
        \begin{subfigure}[b]{0.329\textwidth}
\includegraphics[width=0.999\textwidth]{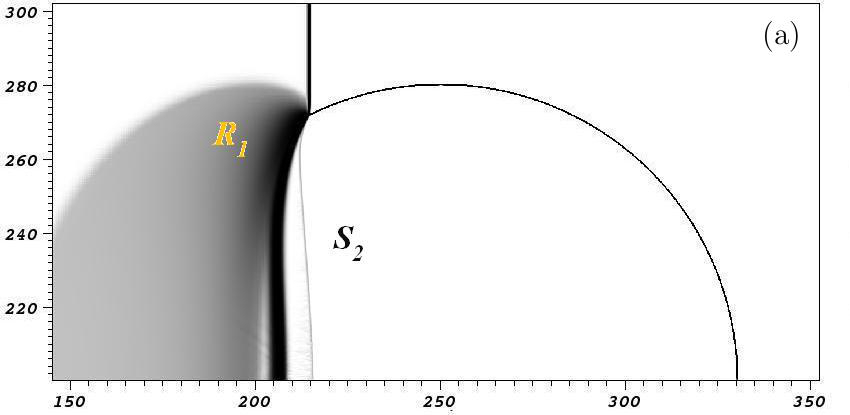}
        \end{subfigure}
        \begin{subfigure}[b]{0.329\textwidth}
\includegraphics[width=0.999\textwidth]{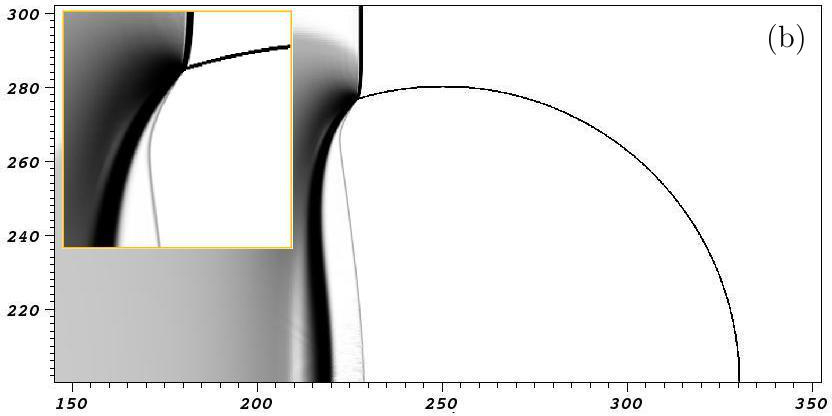}
        \end{subfigure}
\begin{subfigure}[b]{0.329\textwidth}
\includegraphics[width=0.999\textwidth]{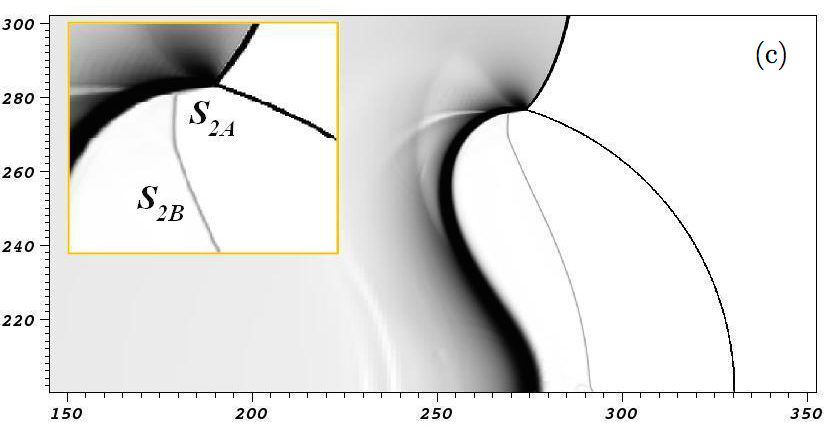}
        \end{subfigure}
\end{minipage}
\begin{minipage}{2\columnwidth}
\centering        
\begin{subfigure}[b]{0.329\textwidth}
\includegraphics[width=0.999\textwidth]{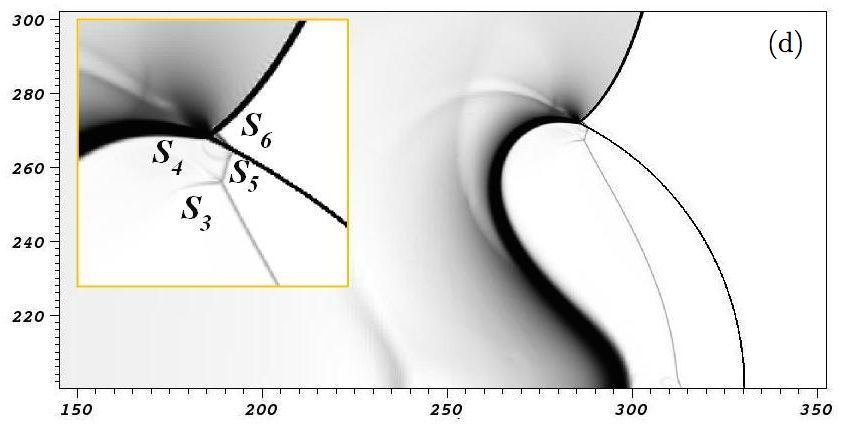}
        \end{subfigure}
        \begin{subfigure}[b]{0.329\textwidth}
\includegraphics[width=0.999\textwidth]{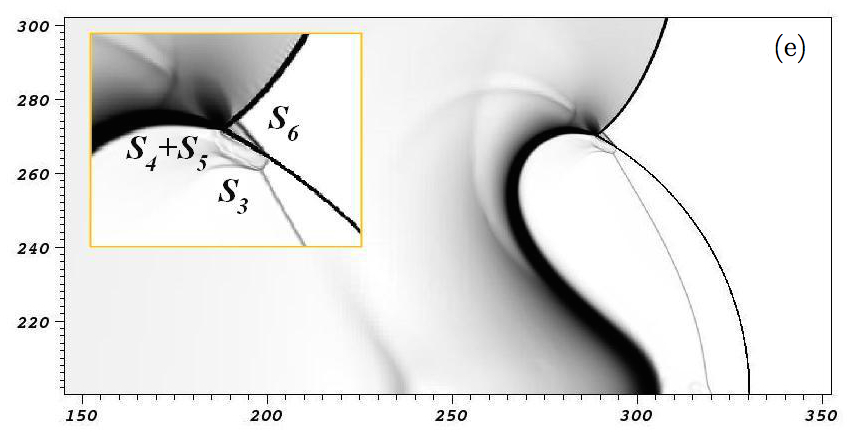}
        \end{subfigure}
\begin{subfigure}[b]{0.329\textwidth}
\includegraphics[width=0.999\textwidth]{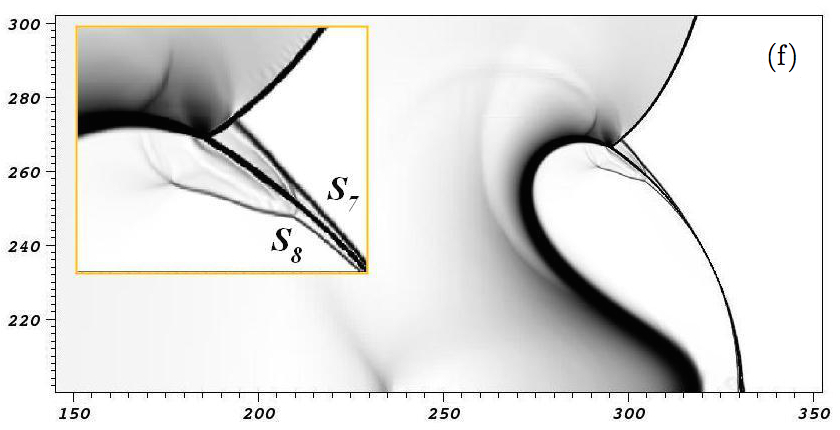}
        \end{subfigure}
\end{minipage}
\begin{minipage}{2\columnwidth}
\centering
        \begin{subfigure}[b]{0.329\textwidth}
\includegraphics[width=0.999\textwidth,scale=0.7]{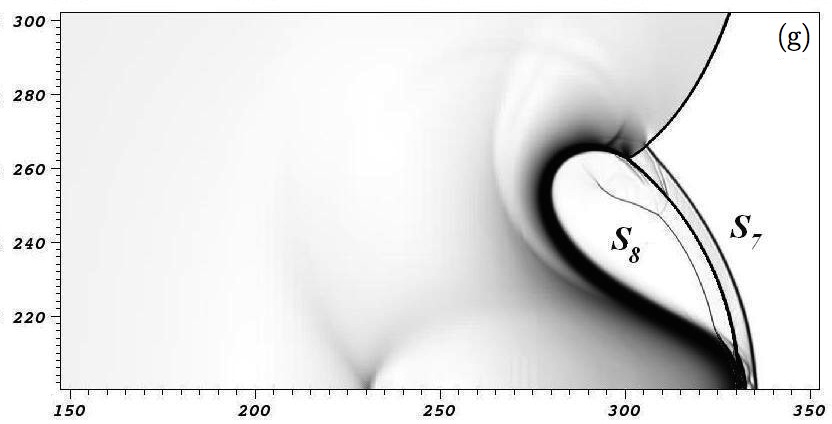}
        \end{subfigure}
        \begin{subfigure}[b]{0.329\textwidth}
\includegraphics[width=0.999\textwidth]{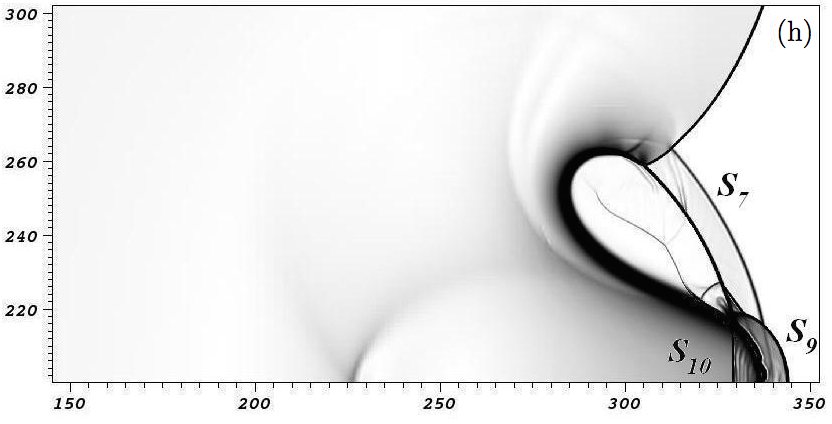}
        \end{subfigure}
\begin{subfigure}[b]{0.329\textwidth}
\includegraphics[width=0.999\textwidth]{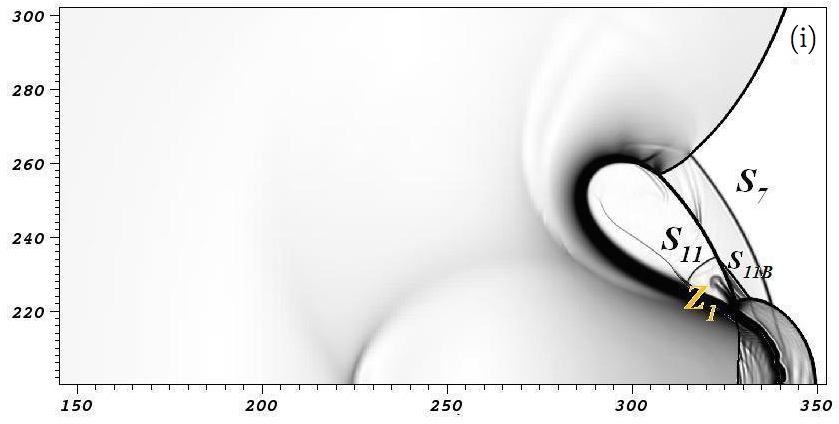}
        \end{subfigure}
\end{minipage}
\caption{Mock-schlieren plots of the early stages in the collapse process
of a single cavity, induced by the impact of a  \SI{10.98}{\giga \pascal}
shock wave, at selected times. The plots
correspond to the $xz$-plane of the three-dimensional field, through the
centre of the cavity. Insets provide magnified regions of the graph, for
a clearer illustration of the weaker waves. Shocks are denoted as `S' and
rarefactions as `R'. The horizontal axis represents $x$  and the vertical
axis $z$, both in \SI{}{\micro \meter}.}
\label{early_single}
\end{figure*}

In order to study the complex wave pattern emanating from the cavity collapse
process, $xz$-slices through the centre of the cavity are taken. The $xy$-plane plots show the same results and are thus omitted. As the cavity collapse is symmetric about the $x$-centreline, the lower part of the cavity is omitted. However, to give the three-dimensional perspective of the setup, we first present in Fig.\ \ref{3Dtemp} a couple of indicative three-dimensional plots of the shock-cavity collapse in terms of the nitromethane temperature field.
Fig.\  \ref{early_single} illustrates the mock-schlieren plots on this plane, for the early stages of the collapse process.
The horizontal axis gives the $x$-position  and the vertical axis the $z$-position,
both in \SI{}{\micro \meter}.
 
In Fig.\  \ref{early_single}a, the transmitted air shock ($S_2$) and reflected
rarefaction wave ($R_1$) generated by the interaction of the ISW (or $S_1$) with the
cavity interface are depicted. In Figs.\ \ref{early_single}b-c, the
non-uniform pressure and velocity fields around the cavity are seen to lead
to the formation of the intruding jet.

Before the jet is fully formed, the air shock $S_2$ traverses a small part 
of the cavity and its upper part moves downwards due to the high pressure
around the upper part of the cavity.  This results in self-focusing of the
shock, leading to a separation of the upper and lower parts of the wave,
labelled $S_{2A}$ and $S_{2B}$  in Fig.\  \ref{early_single}c, and a kink
where the two parts are connected. Two shocks,  $S_3$ and $S_4$,  emanate
from the kink (Fig.\  \ref{early_single}d). The air shock travels faster
than the ISW and the upper part of $S_2$ forms an inward moving shock in
the air, $S_5$, and an outward moving shock in the nitromethane, $S_6$ (Fig.\
 \ref{early_single}d).  Waves $S_4$ and $S_5$ interact inside the cavity,
while a secondary kink and two more shock waves are generated along $S_5$,
as seen in Figs.\  \ref{early_single}e-f. In Fig.\ \ref{early_single}f, the lower part of the air shock ($S_{2B}$) is undergoing a transmission/reflection process\footnote{When a shock wave travelling in a low impedance material reaches a material boundary ahead of which a material of higher impedance is encountered, two new waves are generated; a downstream-travelling shock wave in the high impedance material and an upstream-travelling shock wave in the lower impedance material. In short, we call this a transmission/reflection process.} upon reaching the front cavity face, while the upper part ($S_{2A}$) has already completed this process. The most interesting waves in this case are the ones emanating from the interaction of $S_{2B}$ with the downstream cavity wall. These are the shock labelled $S_7$, travelling downstream in the ambient nitromethane and the shock wave labelled $S_8$ travelling upstream in the cavity (Fig.\  \ref{early_single}g). 

The last row of Fig.\  \ref{early_single} presents the stages close to the collapse time of the cavity. Fig.\  \ref{early_single}g in particular, illustrates the
waves in the system just before the jet reaches the downstream cavity wall.
Upon reaching this wall\footnote{we call this the time of collapse}
two new shock waves are generated in the nitromethane: a downstream-travelling shock wave labelled as $S_9$ and an upstream-travelling shock wave labelled as $S_{10}$, both shown in Fig.\  \ref{early_single}h. Other numerical studies do not distinguish between these two parts of the waves and present it as a single `hammer' shock (e.g.\cite{bourne2002cavity}). However distinguishing between the two will be proven important in the following sections when the temperature field is studied. In the 3D-space, the waves $S_7$, $S_9$ and $S_{10}$ can be considered as spherical caps with different centres, radii and heights. 
Upon collapse, the cavity forms a toroid of trapped air. On each two-dimensional slice this phenomenon is seen as splitting of the cavity in two lobes. Any gas that was in the region between the jet and the downstream cavity wall before the collapse is pushed into the toroid (or lobes), where it is compressed further as time passes. This violent compression results in another shock wave  labelled as $S_{11}$ in Fig.\  \ref{early_single}i travelling upwards in the lobe \cite{ozlem2010numerical} and as this passes over the front cavity wall it generates the shock wave  $S_{11B}$ travelling in nitromethane. 
In the same figure, a liquid infusion zone $Z_1$ is illustrated. This is a zone created by the liquid penetrating into the gas region due to the high pressure acting on the liquid in this area and obtains a toroidal shape in 3D-space. As this zone is treated as a mixture zone by the mathematical model, its thermodynamic details are open to question (similarly in \cite{ozlem2010numerical}).

\subsection{Late stages of the three-dimensional cavity collapse}
\label{3dlate}

\begin{figure*}
\begin{minipage}{2\columnwidth}
\centering
        \begin{subfigure}[b]{0.329\textwidth}
\includegraphics[width=\textwidth]{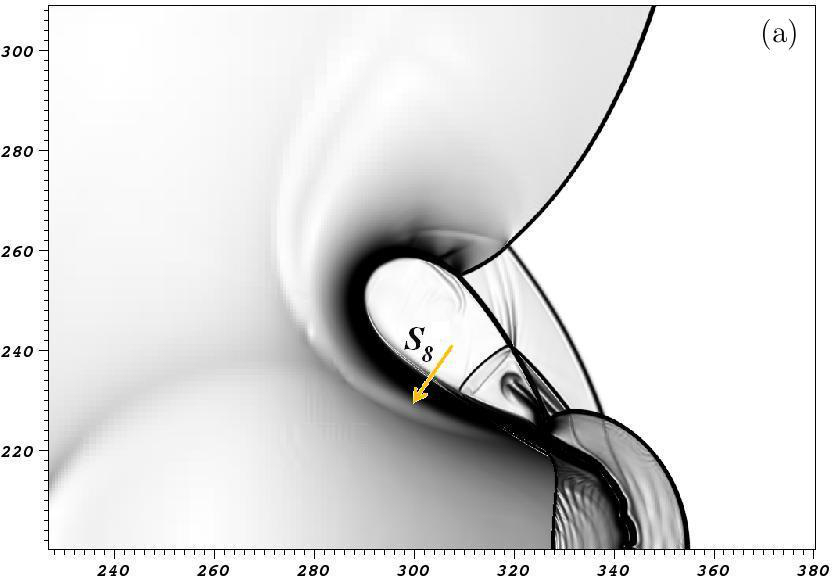}
        \end{subfigure}
        \begin{subfigure}[b]{0.329\textwidth}
\includegraphics[width=\textwidth]{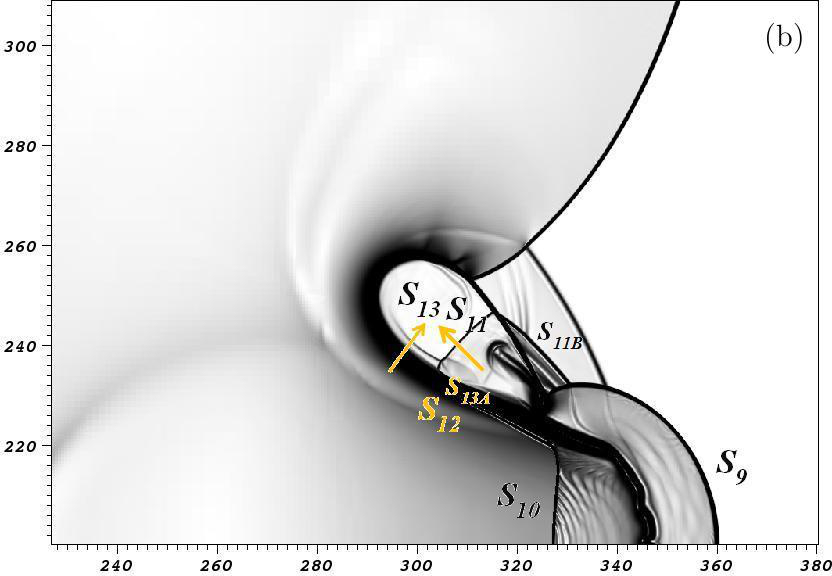}
        \end{subfigure}
\begin{subfigure}[b]{0.329\textwidth}
\includegraphics[width=\textwidth]{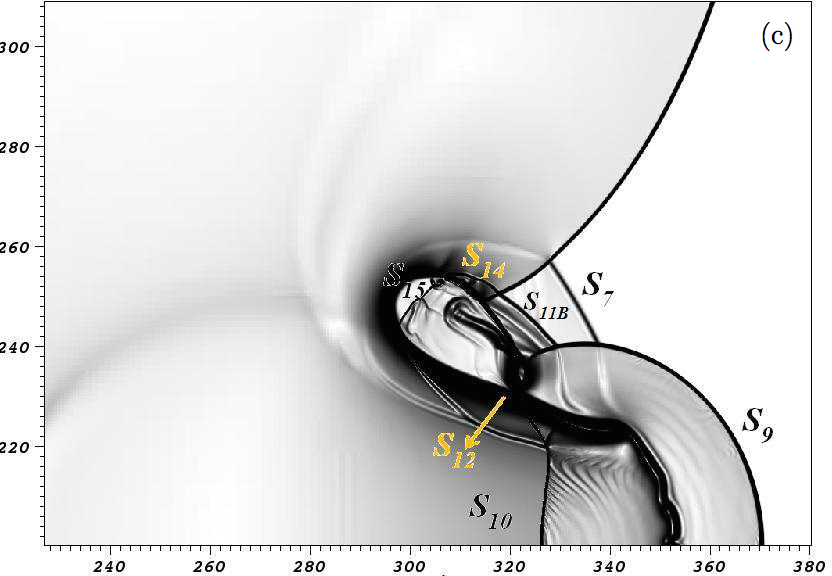}
        \end{subfigure}
\end{minipage}
\begin{minipage}{2\columnwidth}
\centering
\begin{subfigure}[b]{0.329\textwidth}
\includegraphics[width=\textwidth]{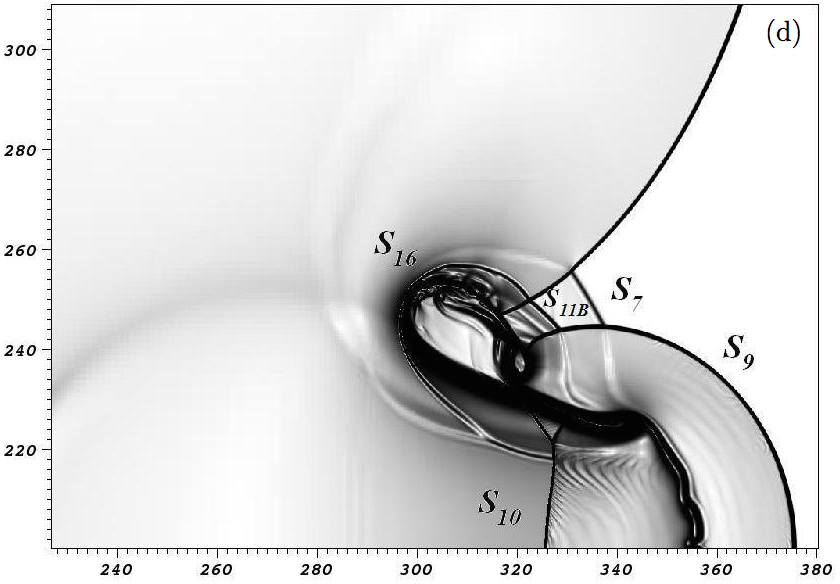}
        \end{subfigure}
        \begin{subfigure}[b]{0.329\textwidth}
\includegraphics[width=\textwidth]{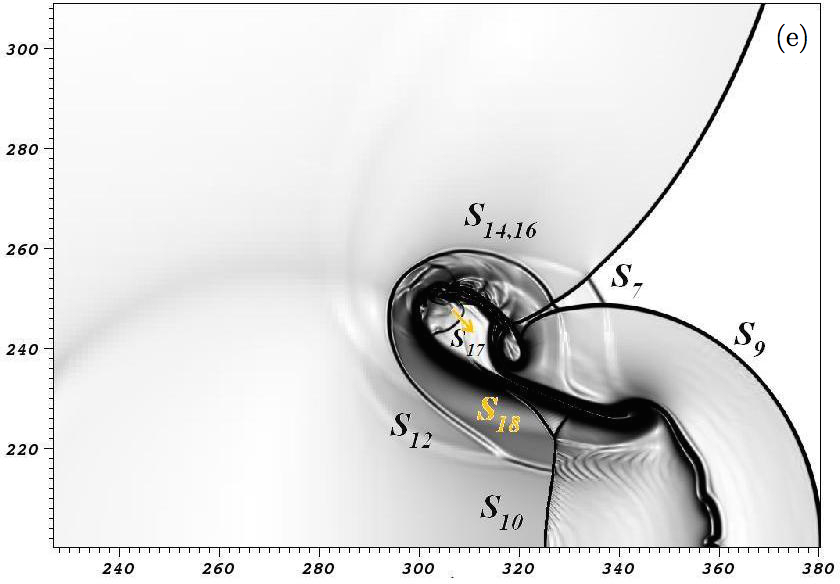}
        \end{subfigure}
\begin{subfigure}[b]{0.329\textwidth}
\includegraphics[width=\textwidth]{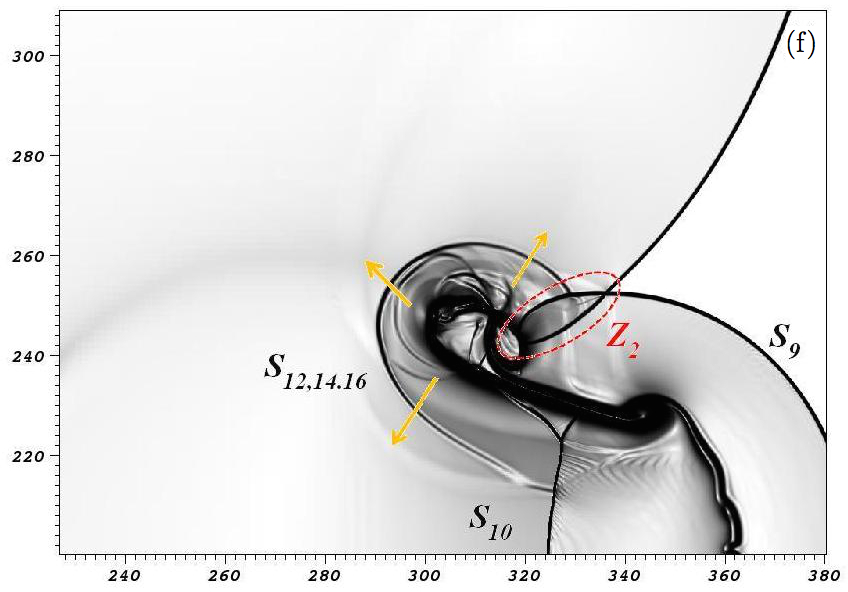}
        \end{subfigure}
\end{minipage}
\caption{Mock-schlieren images of the late stages in the collapse process
of a single cavity, on the central $xz$-plane of the three-dimensional field,
at selected times after the collapse. Shocks are denoted as `S' and
zones as `Z'. The horizontal axis represents $x$  and the vertical
axis $z$, both in \SI{}{\micro \meter}.  }
\label{lateA_single}
\end{figure*}
\begin{figure*}[!t]
\begin{minipage}{2\columnwidth}
\centering
        \begin{subfigure}[b]{0.329\textwidth}
\includegraphics[width=\textwidth]{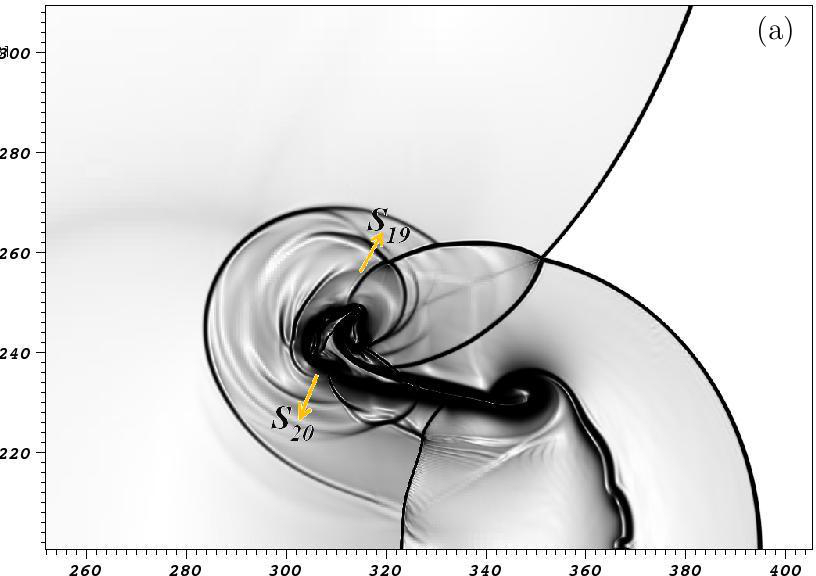}
        \end{subfigure}
        \begin{subfigure}[b]{0.329\textwidth}
\includegraphics[width=\textwidth]{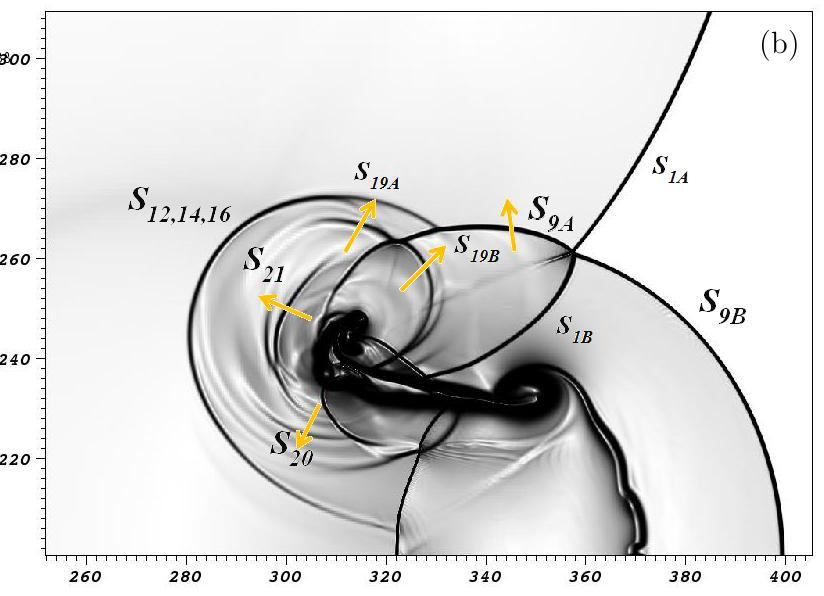}
        \end{subfigure}
\begin{subfigure}[b]{0.329\textwidth}
\includegraphics[width=\textwidth]{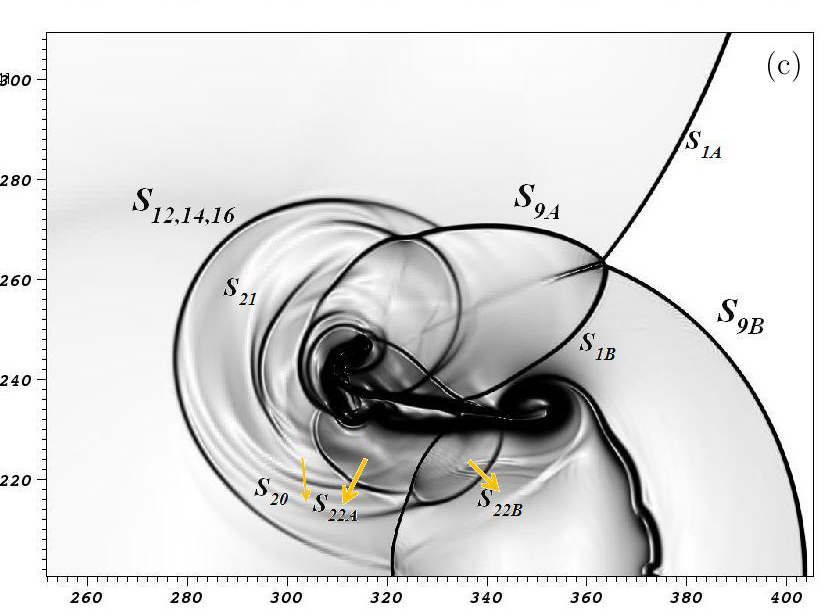}
        \end{subfigure}
\end{minipage}
\begin{minipage}{2\columnwidth}
\centering
\begin{subfigure}[b]{0.329\textwidth}
\includegraphics[width=\textwidth]{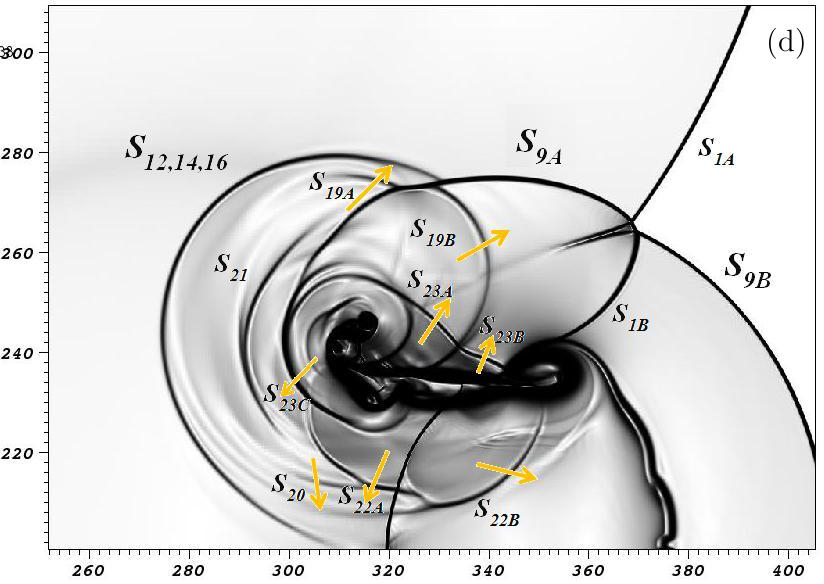}
        \end{subfigure}
        \begin{subfigure}[b]{0.329\textwidth}
\includegraphics[width=\textwidth]{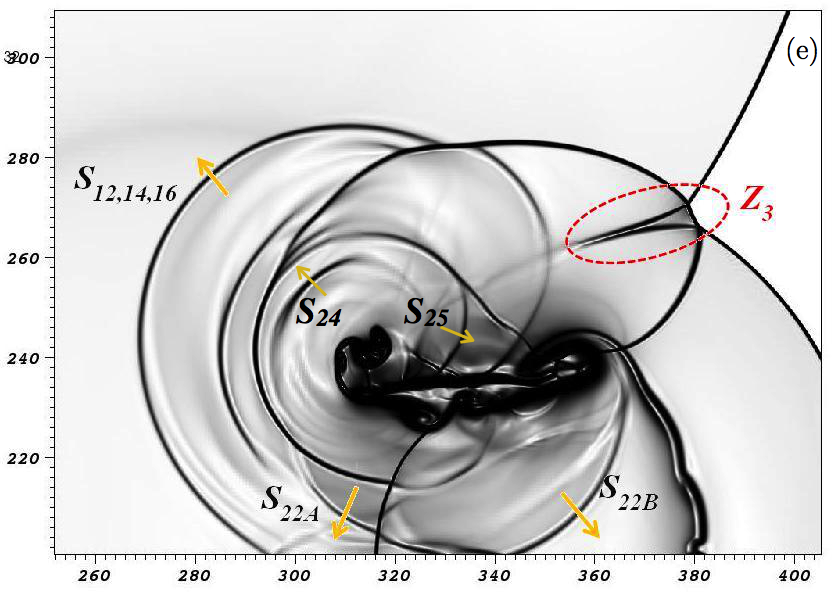}
        \end{subfigure}
\begin{subfigure}[b]{0.329\textwidth}
\includegraphics[width=\textwidth]{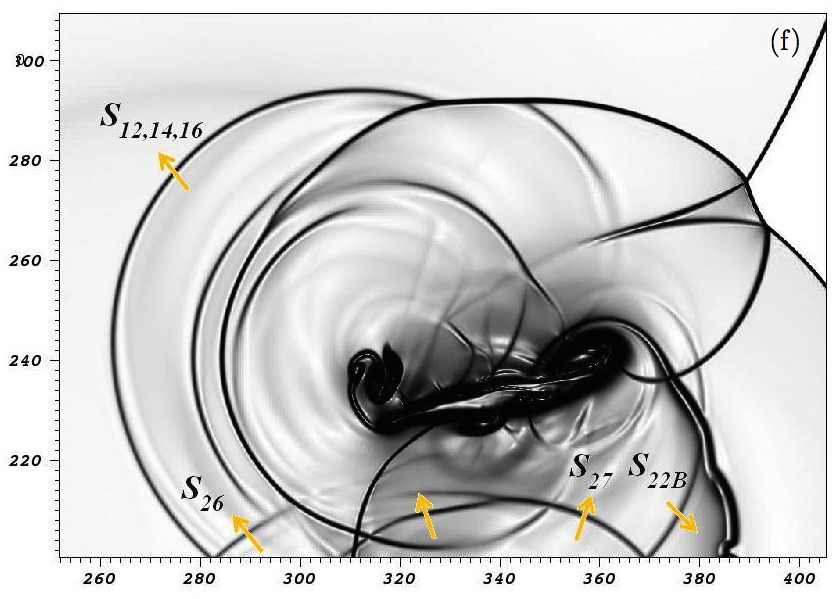}
        \end{subfigure}
\end{minipage}
\caption{Mock-schlieren plots of the late stages (ctd.\ from Fig.\ \ref{lateB_single}) in the collapse process
of a single cavity,  on the central $xz$-plane of the three-dimensional field,
at selected times after the collapse.   Shocks are denoted as `S' and
zones as `Z'. The horizontal axis represents $x$  and the vertical axis $z$,
both  in \SI{}{\micro \meter}. }
\label{lateB_single}
\end{figure*}

 
The wave patterns emerging at the late stages of the collapse process are
 complex, hence  planar mock-schlieren plots remain the most convenient means to visualise them (Figs.\ \ref{lateA_single}-\ref{lateB_single}).
   In Fig.\ \ref{lateA_single}a, the air shock $S_8$ is seen to reach the
downstream side of the lobe. Its interaction with the material interface
results in a shock wave traversing the lobe, $S_{13}$, and a shock travelling in the liquid, $S_{12}$, (see Figs.\ \ref{lateA_single}b-c) obtaining a spherical cap shape in the 3D space. 
The shock wave $S_{11}$ travels upwards in the lobe and interacts obliquely with wave $S_{13}$, resulting in the appearance
of a new pair of weak shock waves, labelled  $S_{13A}$ in Fig.\ \ref{lateA_single}b. As $S_{11}$ continues its upward travel generating $S_{11B}$ it also interacts with the ISW ($S_1$) and this complex process leads to the formation of $S_{14}$. While moving upwards $S_{11}$ also interacts with $S_{13}$ and other weak waves that are present in the lobe forming a front labelled  $S_{15}$. Upon reaching the end of the lobe, it undergoes a transition/reflection process again resulting in a spherical-cap wave ($S_{16}$) in the liquid and a different spherical-cap wave, $S_{17}$,
(Fig.\  \ref{lateA_single}e) in the lobe. The transmitted wave $S_{16}$ 
combines quickly with $S_{14}$ and the two move together outwards from
the lobe, thus labelled as $S_{14,16}$ in Fig.\  \ref{lateA_single}e. This wave then combines
with $S_{12}$, as seen in  Fig.\  \ref{lateA_single}f and is thus labelled as $S_{12,14,16}$. Meanwhile, several
reflection/transmission processes take place at the lobe-nitromethane interface,
which lead to the lobe closing up due to its rapid compression. As a result,
such processes  are difficult to    track from this point
onward  (e.g.\ wave $S_{18}$ in Fig.\  \ref{lateA_single}e). In general,
the waves transmitted from the lobe continue to expand outwards and the ones
traversing the lobe compress the cavity gas further.

The spherical-cap nature of the waves $S_{12,14,16}$ and $S_9$,
together with the curvature of the ISW, lead to the superposition of these
three waves with $S_7$ and $S_{11B}$ in zone $Z_2$, as can be seen in Fig.\
 \ref{lateA_single}f. 

Fig.\  \ref{lateB_single}a illustrates waves $S_{19}$ and $S_{20}$, which
have been generated during late-time transmission/reflection processes at the lobe
boundary. The downstream-travelling wave generated upon the collapse ($S_9$) interacts with the ISW and changes shape, thus labelled now as $S_{9A}$ and $S_{9B}$ (Fig.\  \ref{lateB_single}b). $S_{19}$ interacts with $S_{9A}$ and it is similarly split in two parts, $S_{19A}$ and $S_{19B}$, as illustrated in Fig.\  \ref{lateB_single}b.
New waves, $S_{21}$ and $S_{22},$ are generated by more transmission/reflection
processes at the lobe boundary and $S_{22}$ is also split into two parts, $S_{22A}$
and $S_{22B}$, due to its interaction with the upstream-travelling wave generated upon cavity collapse, namely
$S_{10}$ (Fig.\  \ref{lateB_single}c).
 Waves $S_{23A}$ and $S_{23B}$ are generated by the interaction of the lower part of the ISW ($S_{1B}$)
with the lobe. At this point, the ISW has already been split in two parts
($S_{1A}$ and $S_{1B}$) by its interaction with $S_9$ (Figs.\ 
\ref{lateB_single}c,d). $S_{23C}$ emerges from the lower side
of the lobe and merges with $S_{23A}$ and $S_{23B}$. In Fig.\ $\ref{lateB_single}$f
two new waves are seen to emanate from the lobe, labelled $S_{24}$ and $S_{25}$.
As the ISW continues to interacts with $S_{9A}$ and $S_{9B}$, a Mach stem region
$Z_3$  is formed (Fig.\  \ref{lateB_single}e). This new region, often omitted in the literature, will be seen in the following sections to play an important role in the potential ignition of nitromethane.

In Fig.\  \ref{lateB_single}f wave $S_{26}$ is the equivalent of the $S_{12,14,16}$ wave emanating from the lower lobe of the
cavity (which is not shown here). Similarly, wave $S_{27}$ is the equivalent of waves $S_{22A}$ and
$S_{22B}$ generated at the lower lobe.

\subsection{Temperature distribution during the early stages of the collapse}
\begin{figure*}[!t]
    \centering
\includegraphics[width=0.99\textwidth]{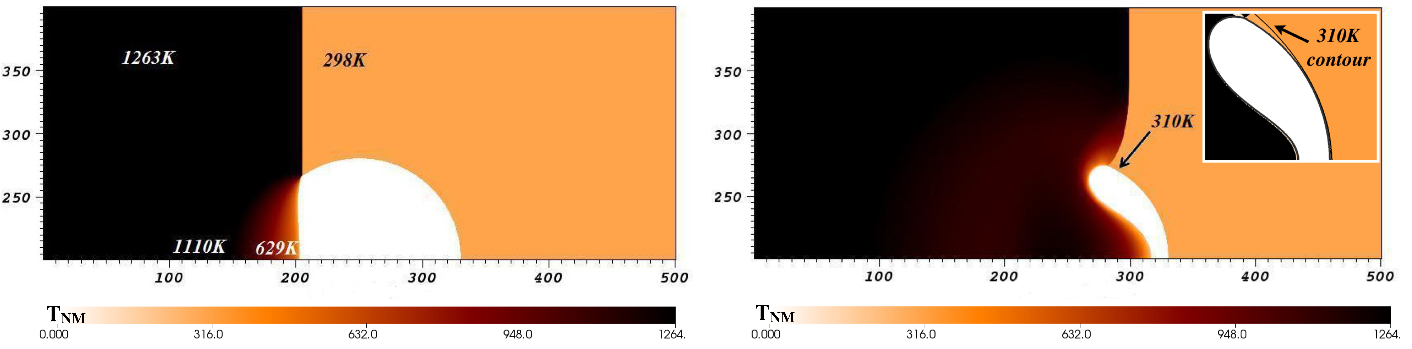}
\caption{Temperature distribution in nitromethane during early stages of
collapse, on the $xz$-slice through the centre of the cavity, corresponding to the three-dimensional plots of Fig.\ \ref{3Dtemp}.}
\label{earlyTNM_single}
\end{figure*}

Having studied in detail the generation of the several waves in the collapse process, we will look now into the effect they have on the nitromethane temperature field. 
When nitromethane is shocked to \SI{10.98}{\giga \pascal}  it attains a temperature
 of  \SI{1263}{\kelvin}, as illustrated in Fig.\  \ref{earlyTNM_single}a. 
Note that since we are only interested in the nitromethane temperature and to ease
the interpretation of the results, the cavity region is masked out  in the
temperature plots of this work. In the early stages of the collapse, the
 rarefaction wave $R_1$ generated by the interaction of the ISW with the
cavity boundary (see Fig.\  \ref{early_single}a), lowers  the temperature
of nitromethane upstream of the cavity. Specifically, the temperature is
distributed according to the distribution of density in the rarefaction wave.
The air shock transmitted into the liquid,  $S_7$ (Fig.\  \ref{early_single}g),
heats the traversed nitromethane region slightly, raising its temperature
from \SI{298}{\kelvin} to about \SI{310}{\kelvin} seen in the inlet of Fig.\  \ref{earlyTNM_single}b.

\begin{figure*}[t]
    \centering
\includegraphics[width=0.99\textwidth]{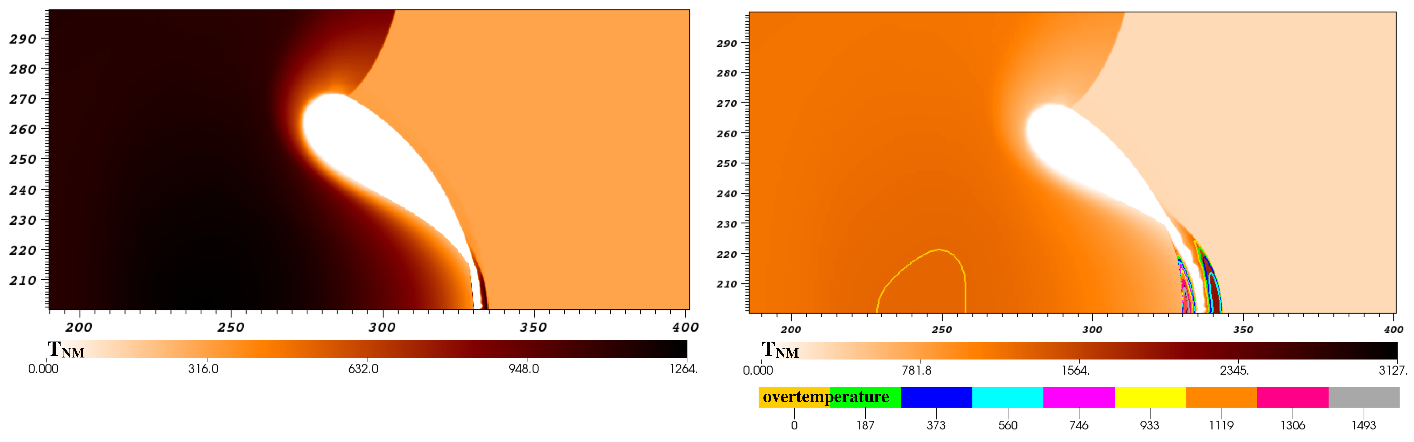}
\caption{Temperature distribution of nitromethane on the $xz$-plane through
the centre of the cavity, close to collapse time. The contours represent
isotherms of over-temperature, which have are meaningful only after the
cavity collapses. The horizontal axis represents
$x$  and the vertical axis $z$, both  in \SI{}{\micro \meter}.}
\label{collapseTNM_single}
\end{figure*}
Fig.\  \ref{collapseTNM_single} shows the temperature field in nitromethane
and the  isotherms of over-temperature, during stages close to the collapse
time. Over-temperature is defined as the local nitromethane temperature minus
the nitromethane post-shock temperature:\  $T_{\text{NM}}-$\SI{1263}{\kelvin} in
this case. This definition is in fact a play on the term `over-pressure' often used in the literature (e.g. \cite{kapila2015numerical}).

 Fig.\  \ref{collapseTNM_single}a illustrates the temperature distribution
just before collapse. The highest temperature at this stage (1263K) is found
behind the ISW. The over-temperature contours are not plotted at this stage
as they have negative values. As soon as the cavity collapses, the new shock
waves generated (waves $S_9$ and $S_{10}$ in Fig.\  \ref{early_single}h)
heat the liquid nitromethane above the post-shock temperature, as seen in
Fig.\  \ref{collapseTNM_single}b. Waves $S_9$ and $S_{10}$ initially generate
over-temperatures of the range \SI{0}{\kelvin}--\SI{400}{\kelvin} downstream
and  of the range \SI{0}{\kelvin}--\SI{1700}{\kelvin} upstream of the cavity,
in regions we will call front hot-spot (FHS) and back hot spot (BHS). In a similar manner we label
the waves generating these hot spots; we call the wave $S_9$ front collapse shock wave (FCSW) as it is the wave generating the FHS and the wave $S_{10}$ back collapse shock wave (BCSW) as it is the wave generating the BHS.
As the BCSW travels outwards, it traverses a region of nitromethane that has already
been shocked by the ISW and as a result, it leads to a generally higher temperature
in the BHS than  the temperature  in the FHS, which is a region that has
not already  been compressed.  As shown in  Fig.\  \ref{collapseTNM_single}c,
over-temperature of the range $0-$\SI{1897}{\kelvin} is observed upstream
of the cavity, in the BHS, while over-temperature of the range $0-$\SI{723}{\kelvin}
 is observed downstream of the cavity, in the FHS. The inner contours are
the ones with the higher over-temperature value but enclose smaller areas.

\begin{figure*}[!ht]
\centering   
\includegraphics[width=0.99\textwidth]{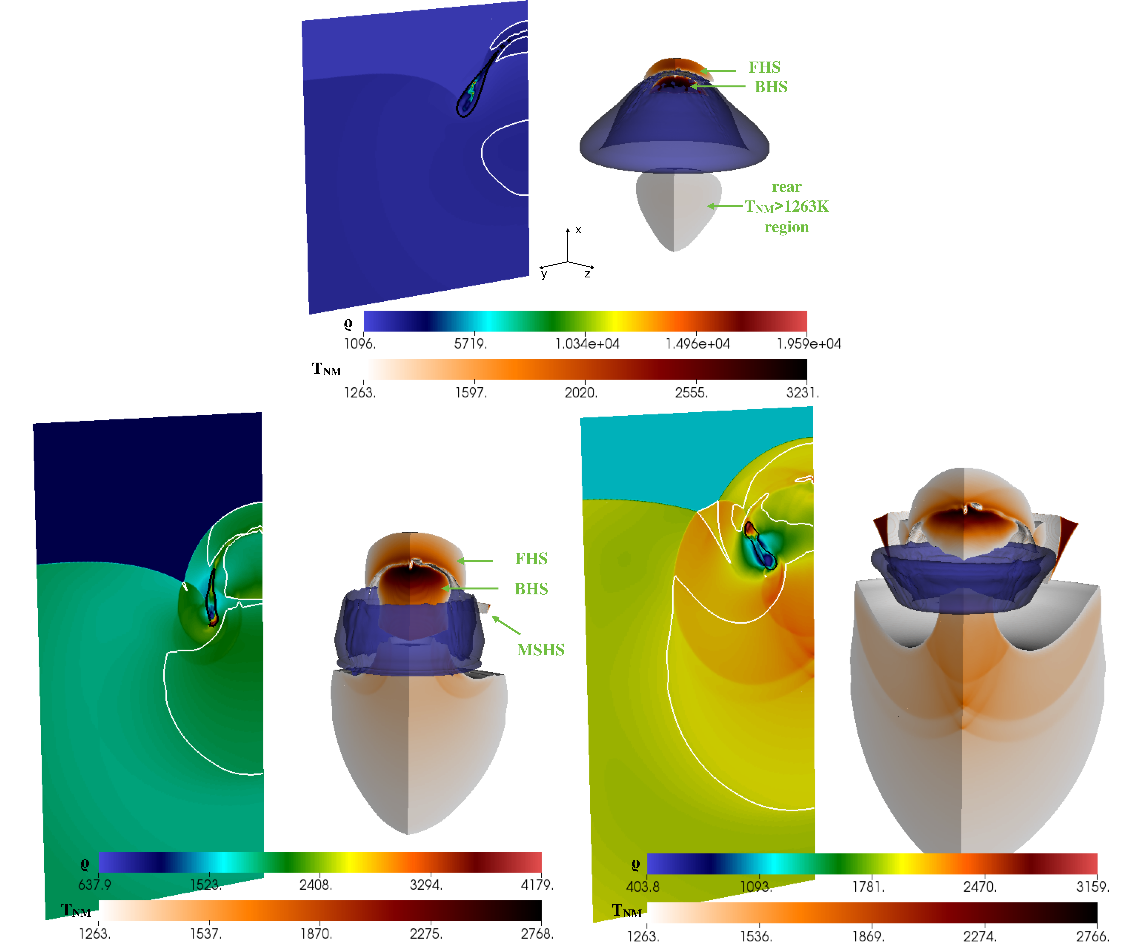}
\caption{Demonstration of over-temperature regions in 3D space. On the left side of each image is a colour plot of the total density field ($\rho$) on a plane through the centre of the cavity and projected backwards for illustration purposes. On each 2D plot, the black contour represents the cavity boundary ($z=0.5$) and the white contours ($T_{NM}=1264$) enclose the regions of (positive) over-temperature. On the right side of each image the three-dimensional cavity boundary is seen as a blue surface (contour $z=0.5$). The over-temperature regions are demonstrated only in one quadrant to aid visualisation, using an 3D isosurface of $T_{NM}=1264$. Note that the collapse process is shown here to move from bottom to top rather than left to right. The images here can be considered as rotated by 90-degrees counter-clockwise compared to Figs. \ref{early_single}-\ref{collapseTNM_single} and \ref{lateA_TNM_single}-\ref{lateB_TNM_single}. This is done solely for illustration purposes.}
\label{3Dpseudo}
\end{figure*}

To visualise the generation of the high-temperature regions we turn to Fig.\ \ref{3Dpseudo}. On the left side of each image, a colour plot of the density field is seen on a plane through the centre of the cavity. The white contour is the $T_{\text{NM}}=\SI{1264}{\kelvin}$ contour, enclosing the regions with positive over-temperatures. On the right side of each image, a three-dimensional plot of the $z=0.5$ contour is seen in blue, denoting the cavity boundary. A three-dimensional equivalent of the  $T_{\text{NM}}=\SI{1264}{\kelvin}$ contour is seen in a the white-orange-black scale. The three-dimensional surface is cut such that it only covers one quadrant to allow better visualisation. At the early stages that are close to the collapse time, the three-dimensional format of the FHS and BHS are seen in Fig.\ \ref{3Dpseudo}(top). At this stage, both high temperature regions have a spherical-cap form. The flow going into the jet creates another region with positive but low over-temperature seen by the lower grey part of  Fig.\ \ref{3Dpseudo}(top).

\subsection{Temperature distribution during the late stages of the collapse}
\begin{figure*}
\centering
    \centering
\includegraphics[width=0.99\textwidth]{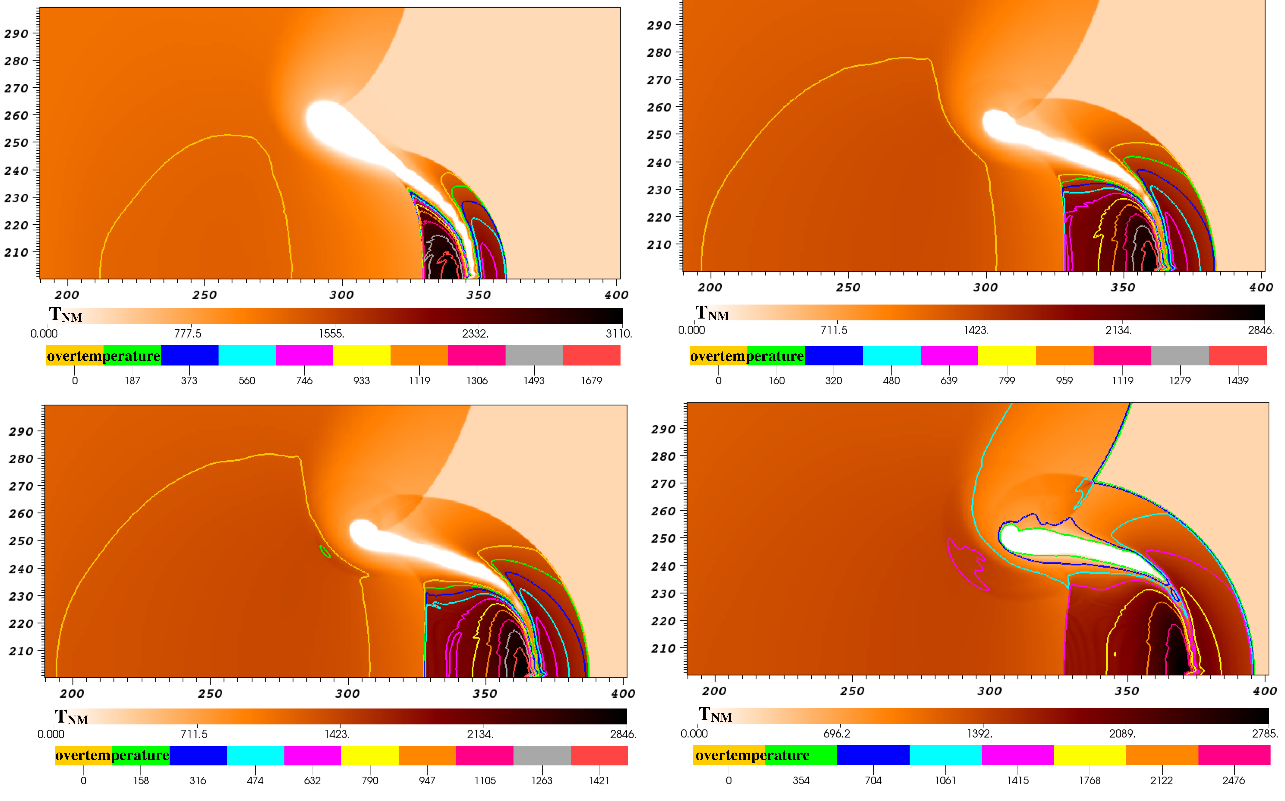}
\caption{Temperature distribution of nitromethane on the central $xz$-plane
 during late stages of the collapse. The
contours represent isotherms of over-temperature. The horizontal axis
represents
$x$  and the vertical axis $z$, both  in \SI{}{\micro \meter}.}
\label{lateA_TNM_single}
\end{figure*}

\begin{figure*}
    \centering
\includegraphics[width=0.99\textwidth]{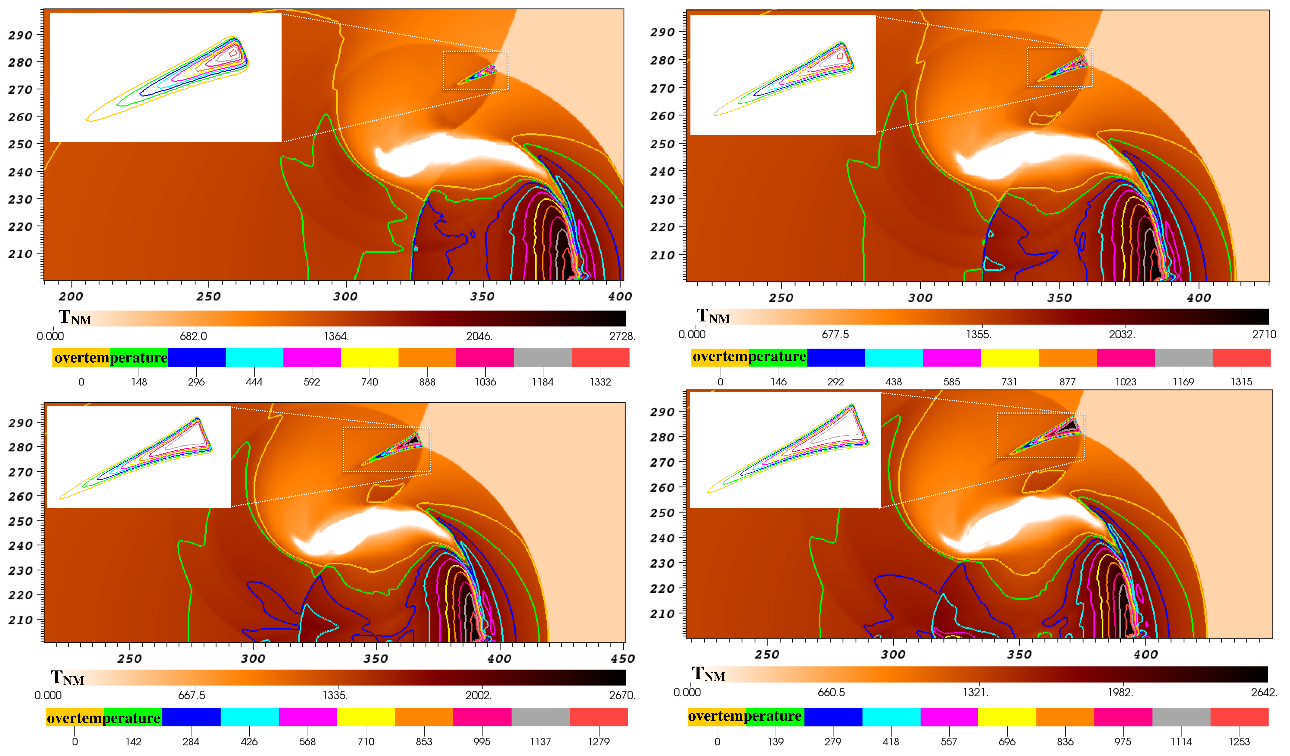}
\caption{Temperature distribution of nitromethane on the central $xz$-plane
during late stages of the collapse. The
contours represent isotherms of over-temperature. The insets provide enlargements
of the Mach stem hot spot region. The horizontal axis
represents
$x$  and the vertical axis $z$, both  in \SI{}{\micro \meter}.}
\label{lateB_TNM_single}
\end{figure*}

The temperature distribution in nitromethane during the late stages of the
collapse process is presented using two-dimensional slices in Fig.\  \ref{lateA_TNM_single} and Fig.\
 \ref{lateB_TNM_single} and three-dimensional contours in Fig.\ \ref{3Dpseudo}.
With the outward propagation of the FCSW and the BCSW($S_9$ and $S_{10}$), the regions of positive over-temperature
grow in volume. The maximum over-temperature, however, decreases. This is
seen to happen eventually in all the over-temperature regions during the late
stages of the collapse. 
The maximum over-temperature in  Fig.\  \ref{lateA_TNM_single}a is \SI{1826}{\kelvin}
 and it is observed in the BHS. The  maximum over-temperature in the FHS
is \SI{823}{\kelvin}. This is contrary to the usual literature reporting of the highest temperature to be ahead of the collapsed cavity (e.g.\ \cite{bourne1992shock,bourne2002cavity,bourne2003temperature}).
The transmitted waves $S_7$ (Fig.\  \ref{early_single}g) and $S_{11B}$ (in Fig.\ \ref{lateA_single}b) do not significantly increase 
the nitromethane temperature. They result in an over-temperature of  only \SI{50}{\kelvin}-\SI{80}{\kelvin}.

In Fig.\  \ref{lateA_TNM_single}a, the temperature around the lobe is seen
to have been affected by the RW. At this point, sample temperatures around
the lobe reach values of about \SI{650}{\kelvin}-\SI{780}{\kelvin}. In Fig.\
\ref{lateA_TNM_single}b the shock wave $S_{12}$ exits the lobe, gets linked
with $S_{14,16}$ and heats the nitromethane as it travels outwards of the
lobe. At this point, sample temperatures in the region traversed by the wave
$S_{12,14,16}$ are in the range of \SI{679}{\kelvin}$-$\SI{745}{\kelvin}.

Meanwhile, region $Z_2$ (Fig.\
 \ref{lateA_single}f) is formed. This zone does  not
play a significant role in increasing the temperature (Fig.\ \ref{lateA_TNM_single}c), yet, as nitromethane is
heated only up to about \SI{760}{\kelvin} in this region. The maximum over-temperature
is \SI{1570}{\kelvin} and is found in the BHS. The maximum over-temperature
in the FHS reaches \SI{750}{\kelvin}.

In Fig.\  \ref{lateA_TNM_single}c, the growth of wave $S_{12,14,16}$ generates
a new over-temperature region upstream of the cavity, part of which is shown
to be bounded by the green over-temperature contour. An isotherm of over-temperature
\SI{0}{\kelvin} enclosing a region of maximum temperature \SI{1345}{\kelvin}
is seen in the same figure.

As more shock waves emanate from the lobe, new regions of over-temperature
are formed around it (see Fig.\  \ref{lateA_TNM_single}d). As the wave
$S_{12,14,16}$ is propagated downwards, it leads to the growth of the over-temperature
regions, like the one illustrated by the fuchsia contour in Fig.\  \ref{lateA_TNM_single}d.
Even though only one contour of this region is included in this figure, the
nitromethane temperature is increased
throughout the whole region traversed by this wave, as inferred by the values
of over-temperature in this figure. 
 
 Meanwhile, the ISW and the FCSW ($S_{1}$ and $S_{9}$) generate zone $Z_3$, which we call  the Mach stem hot spot (MSHS). This is
a high temperature region which, at this instance, encloses over-temperature
contours of values \SI{0}{\kelvin}$-$\SI{230}{\kelvin}, as seen in Fig.\
 \ref{lateA_TNM_single}d. The highest temperature (\SI{2773}{\kelvin}), however,
is still observed within the over-temperature contours generated by the BCSW,
upstream of the cavity. 

However, as zone $Z_3$ grows, the local nitromethane temperature increases,
as  seen in Fig.\  $\ref{lateB_TNM_single}$a and encloses over-temperatures
ranging between \SI{0}{\kelvin} and \SI{1380}{\kelvin}. This is a region that is omitted in previous studies found in the literature (e.g.\ \cite{bourne2002cavity,bourne2003temperature}). Its importance is clear as so high temperatures are found in this region that could lead to ignition in reactive simulations.

The maximum over-temperature (\SI{1421}{\kelvin}) is still located within
the BHS. However, under other initial conditions (e.g.\ weaker ISW) not presented in this work, the authors have observed the maximum
temperature to move in zone $Z_3$ at this stage. The 
highest over-temperature in the FHS is \SI{640}{\kelvin}. 

Fig.\ \ref{3Dpseudo}(bottom left) shows in 3D the high temperature regions of FHS, BHS and the very early stages of the generation of the MSHS. Behind the cavity, a large region of positive over-temperatures is seen with relatively low over-temperature values, though, compared to the other three regions.

Waves $S_{12,14,16}$ and $S_{22B}$ now also travel downwards,
enlarging the over-temperature regions, as seen in Fig.\  \ref{lateB_TNM_single}a.
As wave $S_{12,14,16}$ is superimposed with the BCSW ($S_{10}$) it generates the over-temperature
regions enclosed in dark blue and turquoise, with over-temperature values
\SI{361}{\kelvin} and \SI{446}{\kelvin} respectively. The spherical nature
of waves $S_{12,14,16}$ and $S_{22B}$ leads to the growth of the over-temperature
regions generated in a three-dimensional sense, all around the lobe. 

Wave $S_{26}$ is generated from the reflection of the wave $S_{12,14,16}$
at the bottom boundary (Fig.\ \ref{lateB_TNM_single}b). As a result, the
liquid in the region the wave $S_{26}$ traverses is shocked for the third
time, leading to a temperature of  about \SI{1622}{\kelvin}. Part of $S_{26}$
is superimposed with the BCSW, leading to a temperature of about \SI{1748}{\kelvin}.
The highest temperature in the BHS is now \SI{2600}{\kelvin}, in the FHS
\SI{586}{\kelvin} and in the MSHS \SI{2500}{\kelvin}. 

In Fig.\ \ref{lateB_TNM_single}d, the wave $S_{22B}$ is also reflected
at  the bottom boundary, generating wave $S_{27}$, which is then superimposed
with the BCSW and waves $S_{22A}$ and $S_{22B}$. This leads to temperatures
of the range  \SI{1580}{\kelvin}--\SI{1650}{\kelvin}.

\begin{figure}[!t]
\centering
\includegraphics[width=0.45\textwidth]{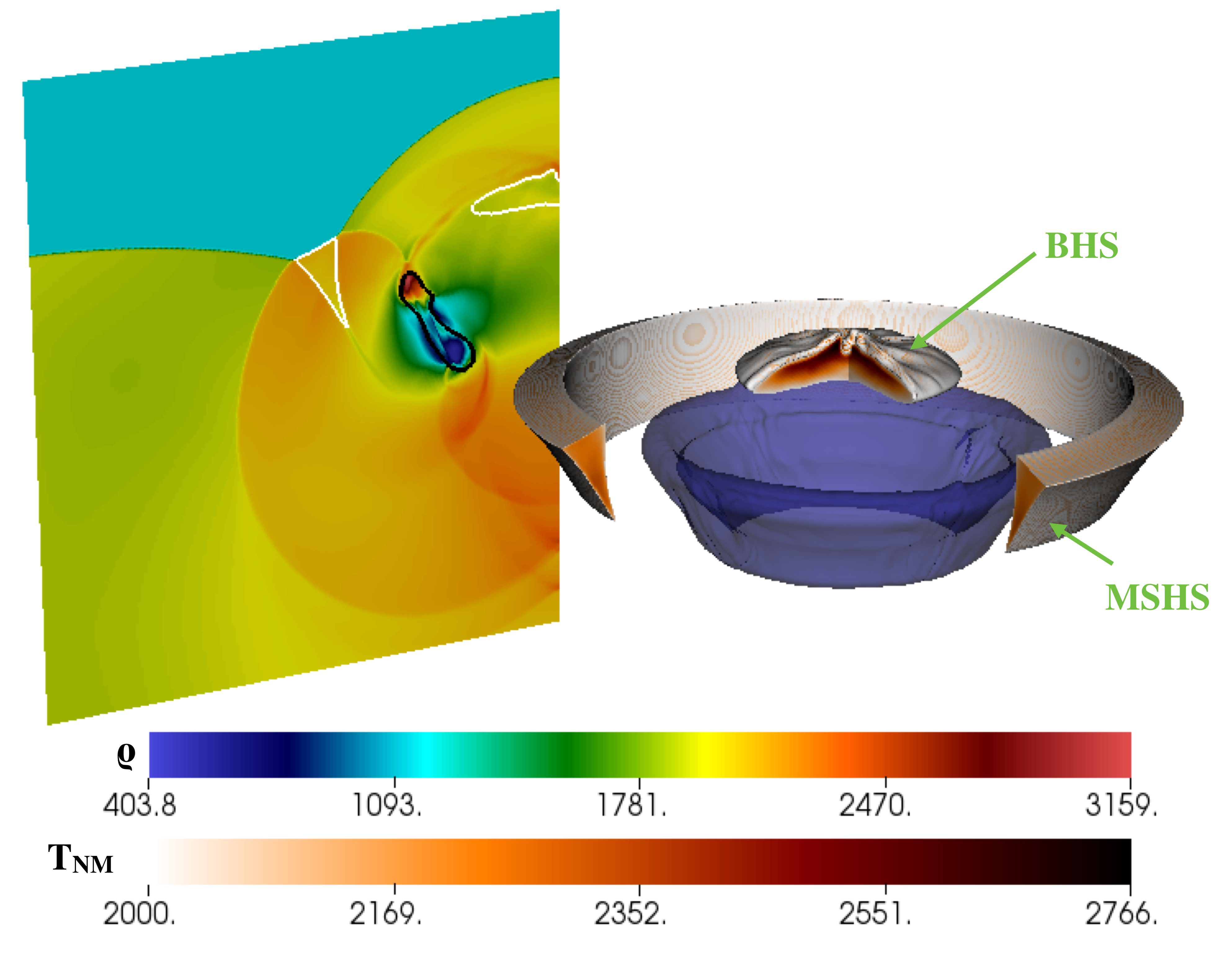}     
\caption{Illustration of the highest late-stage temperature regions in the collapse process. These are found in the back hot spot (BHS) and Mach stem hot spot (MSHS) and are shown here on the right in three-quarters of the 3D domain as isosurfaces of $T_{NM}=2000K$. These are the regions identified as the most probable to observe local ignition at.  The three-dimensional cavity boundary is seen as a blue surface (contour $z=0.5$). On the left side is a colour plot of the total density field ($\rho$) on a plane through the centre of the cavity and projected backwards for illustration purposes. The black contour represents the cavity boundary ($z=0.5$) and the white contours ($T_{NM}=2000K$) enclose the regions of high over-temperatures. Note that the collapse process is shown here to move from bottom to top rather than left to right. The image here can be considered as rotated by 90-degrees counter-clockwise compared to Figs. \ref{early_single}-\ref{collapseTNM_single} and \ref{lateA_TNM_single}-\ref{lateB_TNM_single}. This is done solely for illustration purposes. }
\label{3Dpseudoextra}
\end{figure}

Fig.\ \ref{3Dpseudo}(bottom left) shows in 3D the growth of the four high temperature regions and also the additional increase of temperature in the back cloud of small over-temperature due to the traversing and superposition of the waves emanating from the lobes along the cavity centreline. The highest temperatures are noted in the BHS and the MSHS, as seen in Fig.\ \ref{3Dpseudoextra} and are identified as the contours enclosing temperatures above 2000K. These are noted as the most possible regions for observing ignition in reactive simulations. In this figure the three-quarters of the three-dimensional temperature contour is seen.

\section{Evolution on constant latitude lines}

\begin{figure*}[!t]
\begin{minipage}{2\columnwidth}
 \centering
 \begin{subfigure}[b]{0.49\textwidth}
\includegraphics[width=0.99\textwidth]{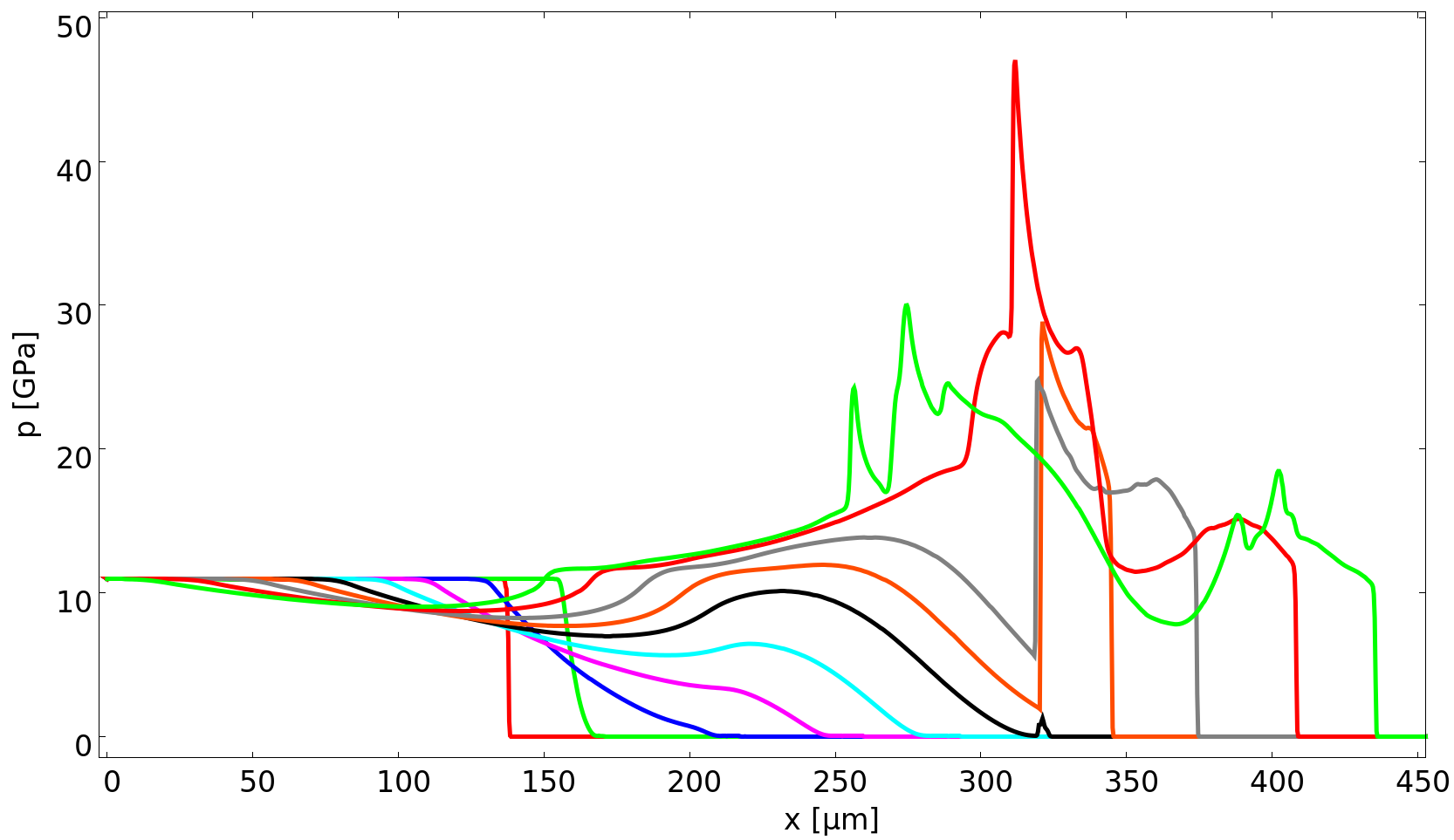}     
\caption{}
\label{lineout200p}
\end{subfigure}
 \begin{subfigure}[b]{0.49\textwidth}
\includegraphics[width=0.99\textwidth]{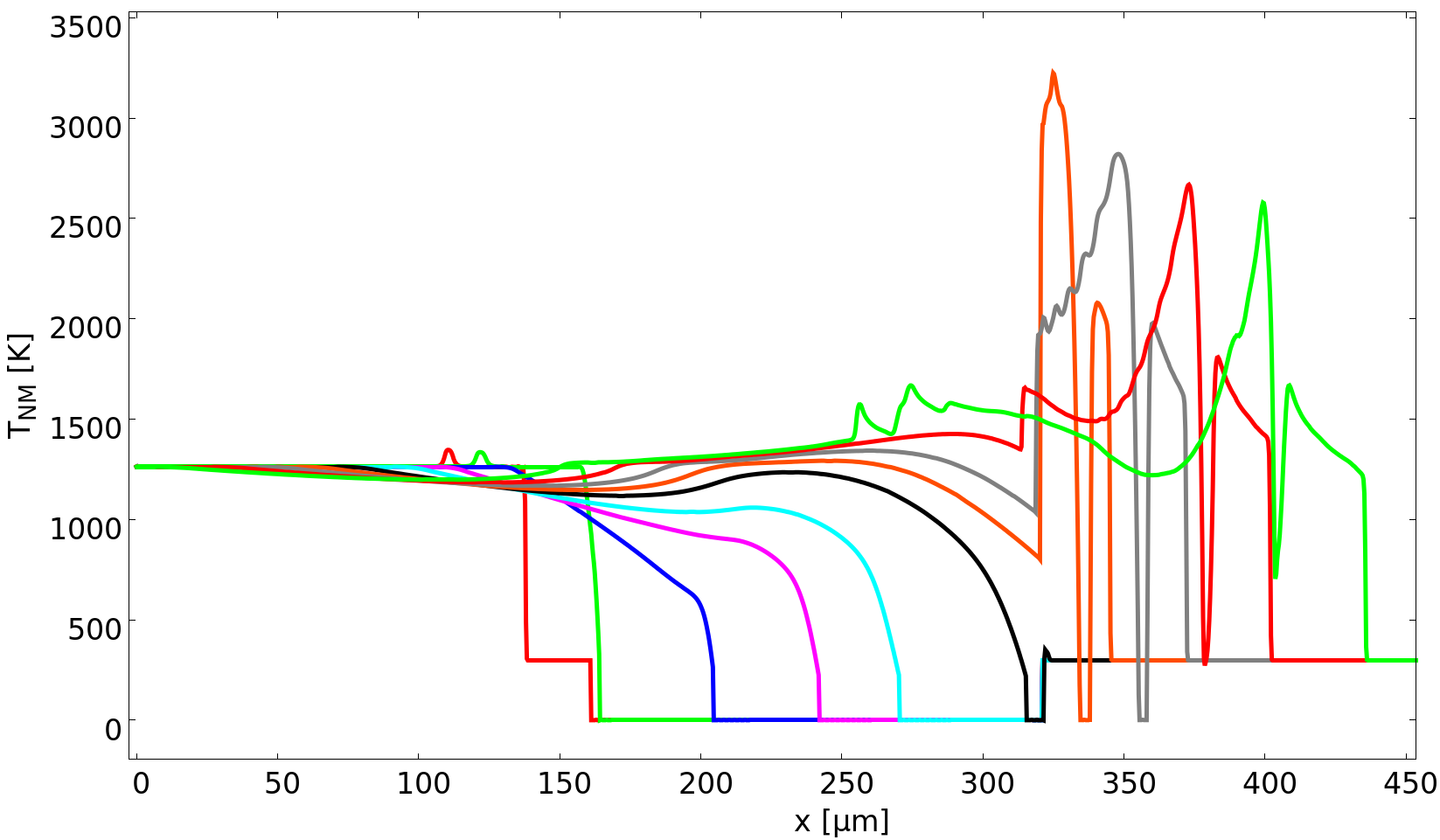}     
\caption{}
\label{lineout200T}
\end{subfigure}
\end{minipage}
\begin{minipage}{2\columnwidth}
    \centering
 \begin{subfigure}[b]{0.49\textwidth}
\includegraphics[width=0.99\textwidth]{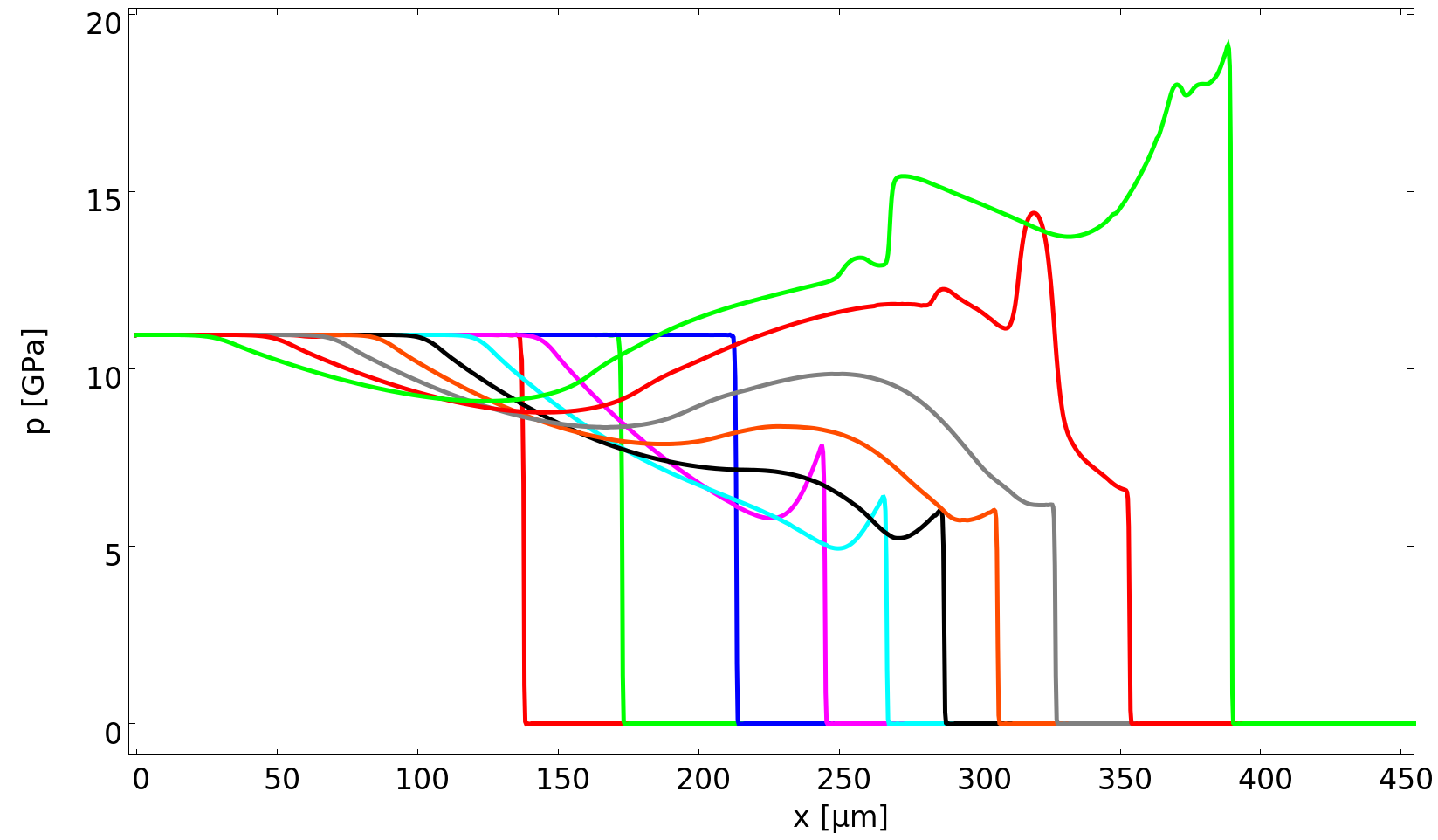}    
\caption{}
\label{lineout290p}
\end{subfigure}
 \begin{subfigure}[b]{0.49\textwidth}
\includegraphics[width=0.99\textwidth]{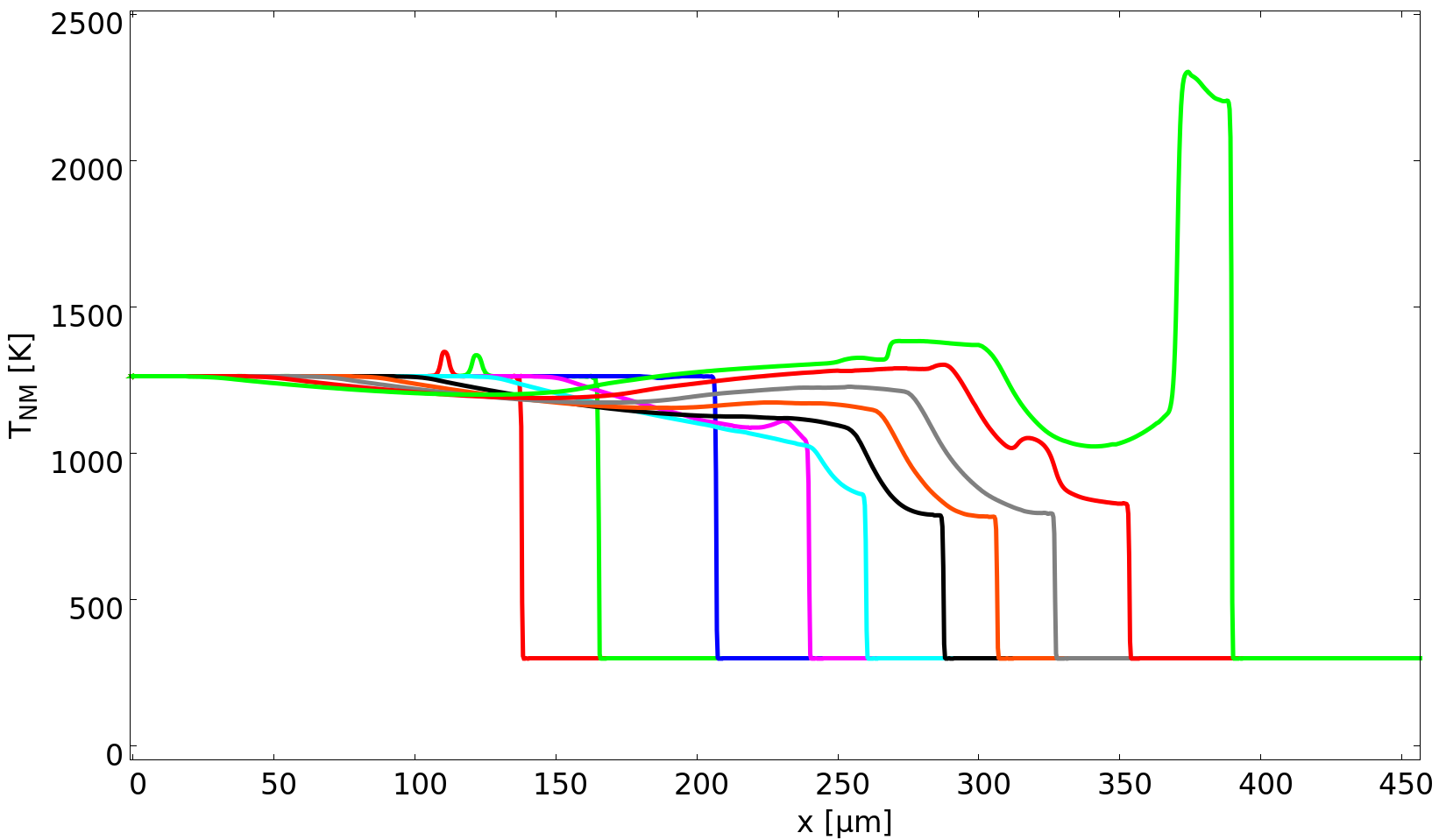}     
\caption{}
\label{lineout290T}
\end{subfigure}
\end{minipage}
\caption{The evolution of the flow field in terms of pressure (left) and nitromethane temperature (right) along the lines   $y=\SI{201}{\micro \meter}$ (top) and   $y=\SI{290}{\micro \meter}$ (bottom).}
\label{lineouts}
\end{figure*}

It is informative to consider the evolution of the flow field on lines of constant latitude on slices like the ones presented in Figs.\ \ref{earlyTNM_single},\ref{collapseTNM_single},\ref{lateA_TNM_single},\ref{lateB_TNM_single}. We consider the pressure and temperature of nitromethane along the lines $y=\SI{201}{\micro \meter}$ (Figs.\ \ref{lineout200p},\ref{lineout200T}) giving insight on the FHS and BHS and $y=\SI{290}{\micro \meter}$  (Figs.\ \ref{lineout290p},\ref{lineout290T}) giving insight on the MSHS. 

In Fig.\ \ref{lineout200p} (first red) the ISW is seen before interacting with the cavity. The corresponding line in Fig.\ \ref{lineout200T} shows the increase of the temperature in nitromethane due to the shock wave to $\SI{1263}{\kelvin}$. Upon the interaction of the ISW with the cavity, the pressure behind the cavity is lowered due to the RW generated (Fig.\ \ref{lineout200p} (blue)), translated into a decrease of nitromethane temperature (Fig.\ \ref{lineout200T} (blue)). As the jet is formed and vortices are generated at the back of the cavity, the RW and the jet flow are working against each other, the former towards lowering the pressure and temperature and the latter towards raising them. The raising of the pressure and temperature prevails as the RW travels upstream while the jet becomes more profound. This is seen in Figs\ \ref{lineout200p} and \ref{lineout200T} (pink-black). 

The FCSW and BCSW  are seen in  Fig.\ \ref{lineout200p} (orange) and correspond to a downstream-travelling and an upstream-travelling thermal front respectively seen in  Fig.\ \ref{lineout200T}. These fronts are the ones expected to lead to ignition in reactive simulations. As the temperature rise from the BCSW is higher than that from the FCSW, we expect the BHS to lead to faster reaction than the FHS. In between the two thermal fronts of \ref{lineout200T} (orange, grey) the nitromethane temperature is seen to reach 0. This is because this region is still occupied by the air of the collapsed cavity hence nitromethane does not reside there. The superposition of waves emanating from the two lobes along the centreline of the cavity is seen as two new, upstream pressure and temperature peaks in the second green graph of Figs.\ \ref{lineout200p} and \ref{lineout200T}.

Figs.\ \ref{lineout290p} and \ref{lineout290T} show the evolution of the pressure and nitromethane temperature along the   $y=\SI{290}{\micro \meter}$ line, at the same time instances as presented in  \ref{lineout200p} and \ref{lineout200T}. As the RW takes longer to reach this line, the pressure and temperature seem to maintain their post-shock values for longer (Figs.\ \ref{lineout290p}, \ref{lineout290T} (first red-blue)). The effect of the RW is seen in the pink and light blue curves. The generation of the Mach stem is seen in the second red and second green lines of  Fig.\ \ref{lineout290p}, accompanied with the thermal fronts shown in Fig.\ \ref{lineout290T}. This would be another region where ignition is expected to be observed in reactive simulations.

\section{Maximum temperature in 2D vs 3D}

\begin{figure}[!t]
\centering
\includegraphics[width=0.45\textwidth]{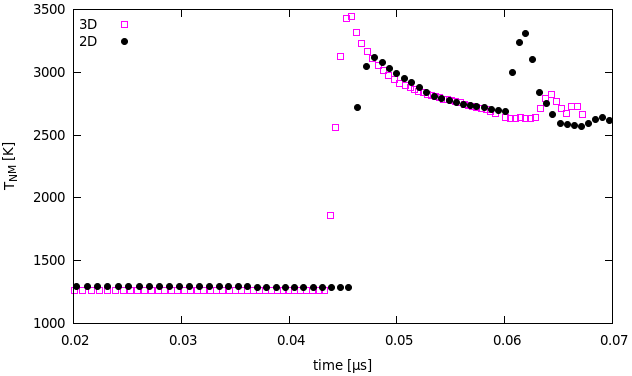}     
\caption{Maximum nitromethane temperature observed in the computational domain (2D or 3D) over time. The black circles denote results from the two-dimensional setup and the pink squares from the three-dimensional set up. }
\label{2Dvs3DTmax}
\end{figure}

To demonstrate the importance of three-dimensional simulations we perform the equivalent two-dimensional simulation of the configuration presented in Sec.\ \ref{3Dresults} and compare directly the temperature fields obtained from the two set ups. In Fig.\ \ref{2Dvs3DTmax} we compare the maximum temperature field observed in the computational domain over time in 2D and 3D simulations.

It is observed that in the three-dimensional simulation the maximum temperature is seen upon collapse. This agrees with the results observed in the detailed analysis performed in Sec.\ \ref{3Dresults} where the maximum temperature is observed in the BHS. The maximum temperature is seen to decrease after this peak, which also agrees with the earlier observation that the temperature reduces as the front and back hot spots grow. A new peak in the maximum temperature is observed at late stages and this corresponds to the temperature increase due to superposition of waves along the centreline of the cavity. In the two-dimensional simulation, the collapse of the cavity is observed slightly later than in the three-dimensional scenario, leading to a lower temperature rise than in the three-dimensional case. However, the decrease of this temperature after the initial collapse peak is almost identical between the two cases. Of great significance is the fact that the maximum temperature in the two-dimensional simulation is observed in the late stages of the collapse.

The maximum temperature results in the two- and three-dimensional configurations follow the same trend, although the first peak temperature is under-predicted and the second peak temperature is over-predicted in the two-dimensional simulations. This would be of great importance in reactive simulation scenarios as, depending on the details of the reaction rate law, it could give the difference of ignition in 3D and no ignition in 2D (at the first peak) or even false late ignition (at the second peak) in 2D. This plot also highlights the importance of the detailed study of the waves and corresponding temperature fields as presented in Sec.\ \ref{3Dresults}, as the maximum temperature can not be entirely accounted for without the detailed wave-pattern analysis.

\section{Conclusions}

In this work we perform resolved numerical simulations of cavity collapse in liquid nitromethane
using a multi-phase formulation, which can accurately recover temperatures in the vicinity of the cavity. Considerable care is taken regarding the form of equations and numerical algorithm to prevent spurious numerical oscillations in the temperature field. The mathematical formulation and its implementation are validated against results of shock-bubble collapse in water studied extensively in the literature, demonstrating a good match between our work and previous studies. The suitability of the model (evolution equations and equations of state) to capture realistic temperatures in the explosive is demonstrated in a second validation problem. The shock temperature of nitromethane at various shock pressures is compared between our work and a combination of analytic and experimental results showing very good agreement.

Following the validation, we study in detail the shock-induced cavity collapse in inert liquid nitromethane in three dimensions.  We follow the interaction of the incident shock wave with the cavity and the generation of waves during the collapse. For ease of interpretation we split the collapse phenomenon in two regimes; early stages and late stages. We follow the events leading to the generation of locally high temperatures and identify the effect each wave has on the nitromethane temperature field, something that would be very difficult to achieve this experimentally. Moreover, numerical studies on this are very limited, as several challenges have to be overcome for this to be possible, as mentioned in the introduction of his work. We also define as over-temperature the difference between the local nitromethane temperature and the post-shock nitromethane temperature. This concept identifies potential regions that could lead to ignition due to cavity collapse that would not ignite in a neat material shocked at by same shock loading. We study the temperature field using plots on two-dimensional slices of the domain, one-dimensional, constant latitude lines on these slices as well as two- and three-dimensional temperature and over-temperature contours.

Commonly in the literature it is suggested that the locus of highest temperature is in front of the cavity, due to the collapse. We demonstrate that upon collapse two shock wave components are generated that lead to two high-temperature loci. The downstream-travelling shock wave (FCSW) leads to the generation of a high temperature region in front of the cavity (FHS) and the upstream-travelling shock wave (BCSW) leads to the generation of a high temperature region behind the cavity (BHS). We identify that, in fact, the highest temperature during the collapse is observed in the BHS behind the cavity. This is attributed to the fact that this region is shocked twice; once by the ISW before the collapse and one by the BCSW after the collapse. In contrary, the FHS region is only shocked once, by the FCSW having similar strength as the BCSW. During the early stages of the collapse, temperatures more than 2.5 times of the post-shock temperature are observed in the BHS. In the FHS temperatures of 1.6 times the post shock temperatures are observed. 

A lot of attention is often given to the shock waves emanating from the lobes due to multiple wave and material interface interactions. It is identified here, though that these waves do not contribute much to the temperature elevation, leading to temperature increases of 0-100K.

Another region identified as a high-temperature locus with the potential to sustain a hot spot in reactive simulations is the Mach stem region generated at late stages after the collapse.  This region is neglected in the literature. However, we demonstrate that the temperatures observed in the MSHS are higher than the temperatures in the FHS and comparable to the temperatures in the BHS. In particular, temperatures of more than two times the post-shock temperatures are seen in the MSHS. 

A final region with positive over-temperatures is identified far upstream, denoted as rear over-temperature region. Initially this region attains temperatures higher than the post-shock temperature due to the jet flow during the collapse. At late times after the collapse, however, this is also the region where the waves emanating from the lobes are superimposed. For the duration of our simulations this superposition leads to temperature increase of $\sim$450K.

The importance of performing three-dimensional simulations compared to two-dimensional ones is demonstrated in the final section. We study the maximum temperature observed over time in the computational domain and we conclude that the two-dimensional scenario leads to could lead to under-predicting the initiation time or predict false late-time ignition.

The identified positive over-temperature regions will be studied in the second part of this work where three-dimensional reactive simulations will be used to clarify which of these in fact lead to ignition. Having studied the hydrodynamic effects of the collapse on the temperature field, in Part II we study the additional effect of chemical reactions in the timescales and lengthscales of the configurations in consideration.

\appendix
\section{Convergence study}
\label{appendixA}
\begin{figure}[!t]
\centering
\includegraphics[width=0.48\textwidth]{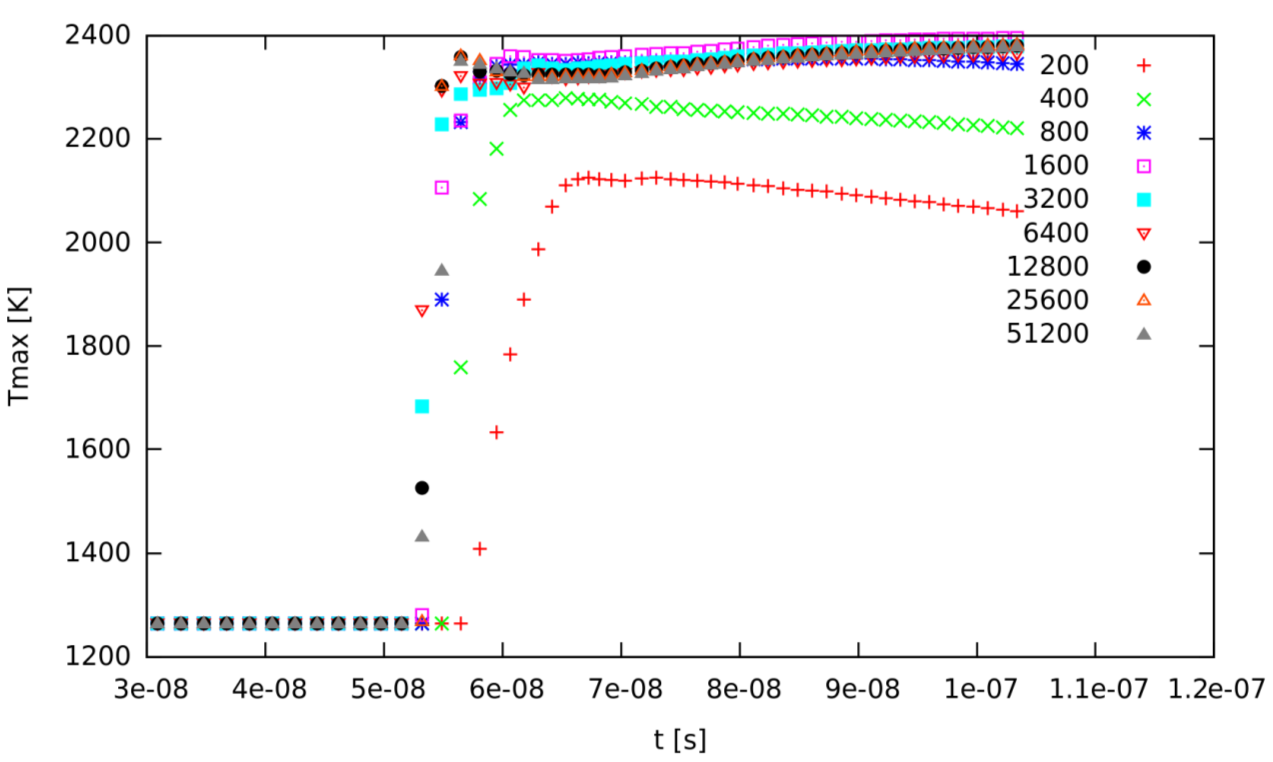}
\caption{Grid-resolution study for the one-dimensional simulation of a cavity
of radius \SI{0.08}{\milli \meter} collapsing in non-reacting nitromethane.
Plot of maximum
temperature \emph{vs} time.}
\label{conv1D}
\end{figure}

To establish the framework for the simulations of this work, grid independence of
one dimensional solutions is assessed for a typical test case of an air cavity
collapsing in nitromethane. The cavity has radius  \SI{0.08}{\milli \meter}
and its centre is initially located at $x=$ \SI{0.25}{\milli \meter},
in a domain of size  \SI{0.5}{\milli \meter}. The initial conditions are as in Table \ref{ICs}.
Nine different resolutions are considered,
starting from 200 cells (64 cells across the cavity) and doubling them each
time until reaching the finest resolution of 512000 cells (16384 cells across
the cavity). In Fig.\  \ref{conv1D}, the effect of grid resolution on the distribution
of the maximum temperature in the domain observed during the collapse is
illustrated. 
 For the 200 and 400-cell meshes, the time of the initial temperature  increase
is not converged. For the 800 and 1600-cell meshes, the time of initial
increase is close to the converged value, but the final temperature plateau
reached is slightly over-estimated.
The resolution of 6400 cells is in very good agreement with the highest resolutions
of 12800 and 51200 cells, for the time of initial increase, the final temperature
plateau and the finer features of the graph. The resolution of 3200 cells
is
in sufficient agreement as well.  

\begin{figure*}
\centering
        \begin{subfigure}[b]{0.24\textwidth}
\includegraphics[width=0.99\textwidth]{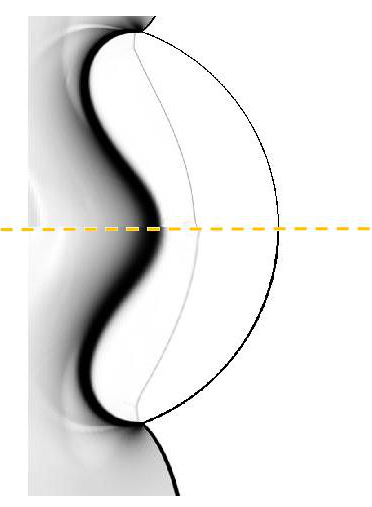}
        \end{subfigure}
        \begin{subfigure}[b]{0.24\textwidth}
 \includegraphics[width=0.85\textwidth]{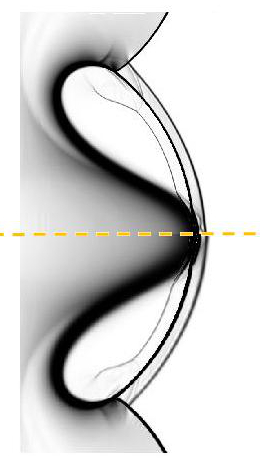}
        \end{subfigure}
\begin{subfigure}[b]{0.24\textwidth}
\includegraphics[width=0.97\textwidth]{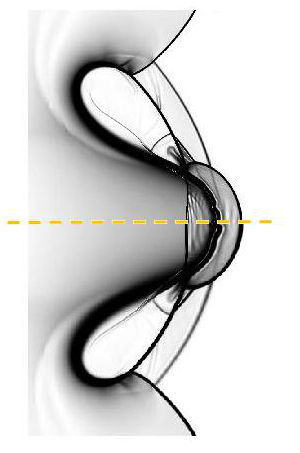}
        \end{subfigure}
\begin{subfigure}[b]{0.24\textwidth}
\includegraphics[width=0.97\textwidth]{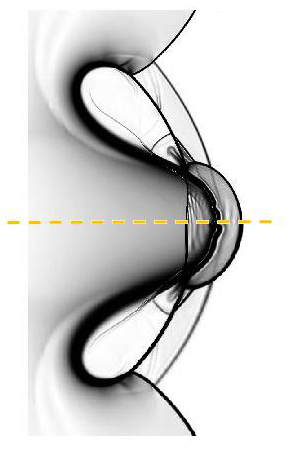}
        \end{subfigure}
\caption{Comparison of the wave pattern at selected times from the two-dimensional shock-cavity collapse in inert nitromethane, performed at different resolutions. The upper part corresponds to a simulation run at a resolution $3200\times1280$ and the lower part to a simulation run at a resolution $1600\times640$.}
\label{conv_sch}
\end{figure*}

The two-dimensional analogue of the one-dimensional cavity collapse simulation
used in the convergence study described above is performed, with $x$-resolutions
of  1600 and 3200 cells.  Only half of the cavity is simulated, with the centreline of the cavity
aligned with the bottom domain boundary in a domain of size  \SI{0.5}{\milli
\meter}$\times$ \SI{0.2}{\milli \meter}. The bottom boundary is reflective,
exploiting the symmetry of the collapse process about the cavity centreline.
In this way the effect of wave patterns generated during the collapse of
the omitted cavity half are taken into account. The $y$-resolution is adjusted
accordingly to 640 and 1280 cells respectively, to obtain square cells. In Fig.\  \ref{conv_sch}
the waves generated during various stages of the two-dimensional simulations
are compared in a mock-schlieren plot between the two grid resolutions. The upper half of each figure
is generated from the $3200\times1280$ simulation while the lower half is
generated from the $1600\times640$ simulation (reflected about its centreline).

 The comparison shows that all the waves captured in the finer-grid simulation
are also captured in the coarser-grid simulation. A comparison between the
corresponding $3200\times1280$ and $6400\times2560$ simulations has shown
that the wave pattern captured by these two simulations is the same. The
same comparison between a $1600\times800$ and a $800\times320$ simulations
revealed that the $800\times320$ simulation is less effective at capturing
the small-scale waves in the lobes and in the ambient nitromethane during
the late stages of the collapse. 

Since the difference in the one-dimensional maximum temperature between resolutions
$1600$ and $3200$ is not significant and the comparison between the two-dimensional
equivalent simulations shows no remarkable difference in the wave pattern,
the choice of grid resolution is made to minimise computational cost. A three-dimensional
simulation of resolution $3200\times1280\times1280$ would require much more
than  \SI{256}{\giga \byte} RAM ($\sim$ \SI{512}{\giga \byte})
 and significantly more computational time than a simulation of resolution
$1600\times640\times640$. The latter requires $\sim$ \SI{96}{\giga \byte} of RAM,
a value that can be reduced for the early stages of the collapse if adaptive
mesh refinement is used effectively.
 In general, without taking into account the effect of AMR, the memory required
for a simulation scales like the cube of the resolution.  As during the late
stages of the collapse the wave pattern is complex, AMR does not reduce
this scaling significantly. Hence, the choice of the $1600$-cell resolution
($\Delta x=$\SI{0.3125}{\micro \meter}) is deemed most appropriate for the
purpose of this work.

\section*{Acknowledgements}
This project was kindly funded by ORICA. The authors also benefited from conversations with A. Minchinton, S.K. Chan
and I.J. Kirby.

\bibliographystyle{spphys}       
\bibliography{shockwaves}   

\end{document}